\documentclass[fleqn,11pt,table]{wlscirep}
\usepackage[utf8]{inputenc}
\usepackage[T1]{fontenc}
\usepackage{times}
\usepackage{amsmath, amssymb}
\usepackage{setspace}
\usepackage{mathtools} 
\usepackage{subcaption}
\usepackage{graphicx}
\usepackage{xcolor}
\usepackage{booktabs}
\usepackage{float}
\usepackage{multirow}
\usepackage[colorlinks=true, allcolors=blue]{hyperref}
\usepackage{listings}
\usepackage{geometry}

\newcommand{\ea}{{\it et al.\ }}

\begin{document}
\title{Breaking the Code: Multi-level Learning in the Eurovision Song Contest}

\author[1,2,3,4]{Luís A. Nunes Amaral*}
\author[4]{Arthur Capozzi}
\author[4,5]{Dirk Helbing} 
\affil[1]{Department of Engineering Sciences and Applied Mathematics, Northwestern University, Evanston, Illinois 60208, USA}
\affil[2]{Department of Physics and Astronomy, Northwestern University, Evanston, Illinois 60208, USA}
\affil[3]{Northwestern Institute on Complex Systems, Northwestern University, Evanston, Illinois 60208, USA}
\affil[4]{Computational Social Science, ETH Z\"urich, Switzerland}
\affil[5]{Complexity Science Hub Vienna, Austria}
\affil[*]{amaral@northwestern.edu}


\doublespacing

\begin{abstract}
Organizations learn from the market, political, and societal responses to their actions. While in some cases both the actions and responses take place in an open manner, in many others, some aspects may be hidden from external observers.  The Eurovision Song Contest offers an interesting example to study organizational level learning at two levels: organizers and participants. 
We find evidence for changes in the rules of the Contest in response to undesired outcomes such as runaway winners.  
We also find strong evidence of participant learning in the characteristics of competing songs over the 70-years of the Contest. English has been adopted as the {\it lingua franca\/} of the competing songs and pop has become the standard genre.  Number of words of lyrics has also grown in response to this collective learning. Remarkably, we find evidence that four participating countries have chosen to ignore the `lesson'  that English lyrics increase winning probability. This choice is consistent with utility functions that award greater value to featuring national language than to winning the Contest. Indeed, we find evidence that some countries --- but not Germany --- appear to be less susceptible to `peer' pressure.   These observations appear to be valid beyond Eurovision.
\end{abstract}

\maketitle

\section*{Introduction}

Contests occur in a multitude of contexts.\cite{van2015theory}
Animal contests determine access to food and mating opportunities\cite{reichert2017cognition}. Sport contests determine popularity, income, and recognition of teams and individuals.\cite{fernandez2016attacking} Product contests determine the access of business firms to markets. \cite{mohr2002managing} Political contest determine access to power by individuals and parties.\cite{de2021political, iyengar2000new, jungherr2016twitter} Contest theory explores how experiences --- i.e., learning opportunities --- inform decision making and the role of evolutionary dynamics.\cite{ribeiro2013move, pastor2009learning} In many of these settings, the rules of the contest can be seen as constant, hiding the possible interplay between contestants' actions and rule setting.  The Eurovision Song Contest (ESC) --- which from now on we will refer to simply as Eurovision or Contest ---  provides a novel and exciting context on which to study such interactions. Scholarly work has demonstrated that the Eurovision is a stage for more than just musical choices (see \cite{garcia2013measuring,dekker2007eurovision,yair1995unite,yair2019douze} and, especially, \cite{yair2022march}). Eurovision involves interactions at multiple organizational levels --- organizers, national selection organizers, competing musicians, and voters.  We will focus here on two of those levels: organizers of the Contest and the participating countries.  The organizers' goal is to maximize television audiences. To achieve this goal, they must ensure that the rules of the Contest are clear and that the outcomes are unpredictable. If the rules are not clear, the audience may feel that the winner was selected in advance. If the outcome is predictable either before the Contest, or even during the voting stage, then the audience will be bored and uninterested.

Participating countries are {\it de facto\/} represented by a public broadcaster, but have wide latitude in determining how they select their competing  songs. For example, the song representing Italy has typically been the winner of the Sanremo Music Festival. But both, artist and song could be selected internally by a committee appointed by the broadcaster.  Typically, there is a clear tension between a country's specific music traditions and the aggregate music taste of voting groups. 

Because the sizes of global and national audiences, competing songs, and voting records are public information, there is ample opportunity for individual and collective learning by all parties.  Here, we investigate these learning processes.  In this, our study differs from others that have usually focused on the characteristics of successful songs, not on interaction and learning effects.

The triumphant winning song of the 2024 Eurovision is well remembered to start with the words ``Welcome to the show. Let everybody know... I broke the code."
While the theme of this song is actually the breaking of gender stereotypes, one might wonder whether they were also suggesting that the band had figured out the formula for success of the Contest itself. Below, we show that some song characteristics can indeed increase the chances of success. However, those rules have not been uncovered by a minority of ingenious songwriters, but were figured out by the participants collectively.

\subsection*{Background}

Already in the past, the study of the dynamics of success has been interested in music songs. For example, Salganik \ea \cite{salganik2006experimental} demonstrated that ranking songs leads to more unpredictable and unequal distributions of success, while the timing and order in which the songs are presented matters. Rosati \ea \cite{rosati2021modelling} showed that the dynamics of song popularity are consistent with a contagious process.  More recently, the wide availability of databases with  information about songs, including metrics of their success, has allowed the development of models that predict song popularity with some reasonable success.\cite{votter2021novel,kamal2021classification,lee2018music,kang2022analyzing}

Importantly, however, most of these studies focus on the success of individual songs, ignoring the context in which those songs were created and presented. Here, we will investigate just this. Success of songs --- and other creative products\cite{wasserman_pnas, wasserman2015correlations} --- is better understood in the context of a specific competitive environment. This is motivated by theories of (co-)evolution \cite{helbing1996stochastic}, where ``fitness'' is not an absolute property, but largely influenced by the other species. Similar approaches have also been applied to the study of innovation dynamics, 
interactive behavioral change \cite{centola2010spread},
and cultural change \cite{creanza2017cultural}. An interesting example for this approach is the ``Art Exhibition Game,"  a behavioral experiment that studied the (co-)evolution of innovation and success in a decision science laboratory \cite{balietti2016peer}.

\section*{Data and Methods}

We retrieved data for all Eurovision Song Contests from its start in 1956 until 2024 from Wikipedia.\cite{wikipedia_eurovision_nodate} For each contest, these pages list the set of competing nations, the performer names, the song titles, the song languages, and the points earned.  The ``points earned'' information is quite detailed, including not only total points in the finals and the semi-finals (from 2004 onward), but also a breakdown of points by awarding nation and by origin (jury or televoting). For most analyses, we focus on a set of 20 ``core" countries which have participated in at least two thirds of the Contest between 1975 and 2000. 

We enriched these data by obtaining song lyrics from the site {\it Letras.com}.\cite{letras_nodate} We then used AI tools to obtain (i) the music genre for each song, (ii) the characteristic words representing the subject matter of a song, and (iii) the audio features for each track. We obtained the translations of the lyrics from the website {\it EurovisionWorld.com}. To quantify the evolution of the themes in song lyrics, we leveraged the 30 song themes taxonomy identified by Henard and Rossetti~\cite{henard2014all}, which can be  associated with distinct sets of keywords. We focus on the 11 themes --- Aspiration, Breakup, Confusion, Desire, Desperation, Escapism, Inspiration, Loss, Nostalgia, Pain, Rebellion --- that appear in over 5\% of all songs. Some of these 11 themes can be seen as relating to painful feelings (Breakup, Confusion), while others may relate to positive states (Aspiration, Desire, Escapism, Inspiration).  Desperation, Pain, and Rebellion may speak of dissatisfaction with the present, while both Loss and Nostalgia may reflect a longing for the past. Recent studies have demonstrated the effectiveness of Large Language Models (LLMs) in topic modeling~\cite{pham-etal-2024-topicgpt, wang2023prompting}, lyric emotion classification~\cite{10825406, REVATHY20231196}, and lyric generation~\cite{lei2024songcreator,10.1145.3474085.3475502}; accordingly, we employed {\it GPT-4o} to annotate each song's lyrics with up to three of the twelve identified themes. More information on this task can be found in the Supplementary Material section.

To characterize the musical characteristics of songs, we employed the API provided by the streaming service Spotify.\cite{spotify_developer} Spotify's API offers thirteen audio features for each track. However, our study focused on six key features that we found to be the most relevant and informative: Acousticness, Danceability, Energy, Loudness, Tempo, and Valence. It’s worth noting that these features are only based on the audio track and do not take the song's lyrics into account. {\it Acousticness\/} is a confidence measure of whether the track is acoustic or not, with values between 0 and 1. {\it Energy\/} is a perceptual measure of intensity and activity that relies on dynamic range, perceived loudness, timbre, onset rate and general entropy. {\it Loudness\/} is the overall volume of a track in decibels (dB). {\it Danceability\/} is based on a combination of musical elements --- including tempo, rhythm stability, beat strength and overall regularity --- relating to whether one can dance easily to the song or not.  {\it Tempo\/} measures the pace at which a section of music is played. {\it Valence\/} describes the musical positivity conveyed by a piece of music. Songs with high valence sound happy and cheerful, while pieces with low valence sound sad or angry.

While Spotify doesn't disclose how exactly its internal system assigns audio features to songs, its API is widely used and considered reliable in research. For example, they’ve been used for the identification of emotions,\cite{panda2021does}, to explain song popularity,\cite{nijkamp2018prediction, gulmatico2022spotipred}, to understand why people listen to music,\cite{duman2022music} and to explore how music streaming behavior relates to the personality traits of Spotify users.\cite{Hongpanarak}

Finally, we categorized the audio tracks according to musical genre. We employed a genre classification model available on Huggingface~\cite{musicgenresclassification}, fine-tuned from the {\it wav2vec2-base-960h\/} model~\cite{baevski2020wav2vec} using the labelled GTZAN Dataset.\cite{sturm2013gtzan} This dataset consists of 1,000 audio samples, each 30 seconds long, evenly distributed across ten music genres: 'Blues', 'Classical', 'Country', 'Disco', 'Hiphop', 'Jazz', 'Metal', 'Pop', 'Reggae', and 'Rock'. After 24 epochs, the model obtained in the training phase had a cross-entropy validation loss of $0.7037$, an area under the ROC curve of $0.964$, and a weighted accuracy of $0.815$.

The model assigns each track a percentage of relevance for one or more of the ten genres listed above. To make our analysis more robust and to better fit these genres to the songs competing, we aggregated the ten genres into three macro categories: {\it Pop\/}, which also includes disco;  {\it Rock\/}, which also includes metal; and {\it Other\/}, which aggregates the remaining 6 genres.

\section*{Results}

\begin{figure}[ht!]
    \centering 
    \begin{tabular}{p{0.48\linewidth} p{0.48\linewidth}}
        \vspace*{-0cm} \includegraphics[width=\linewidth]{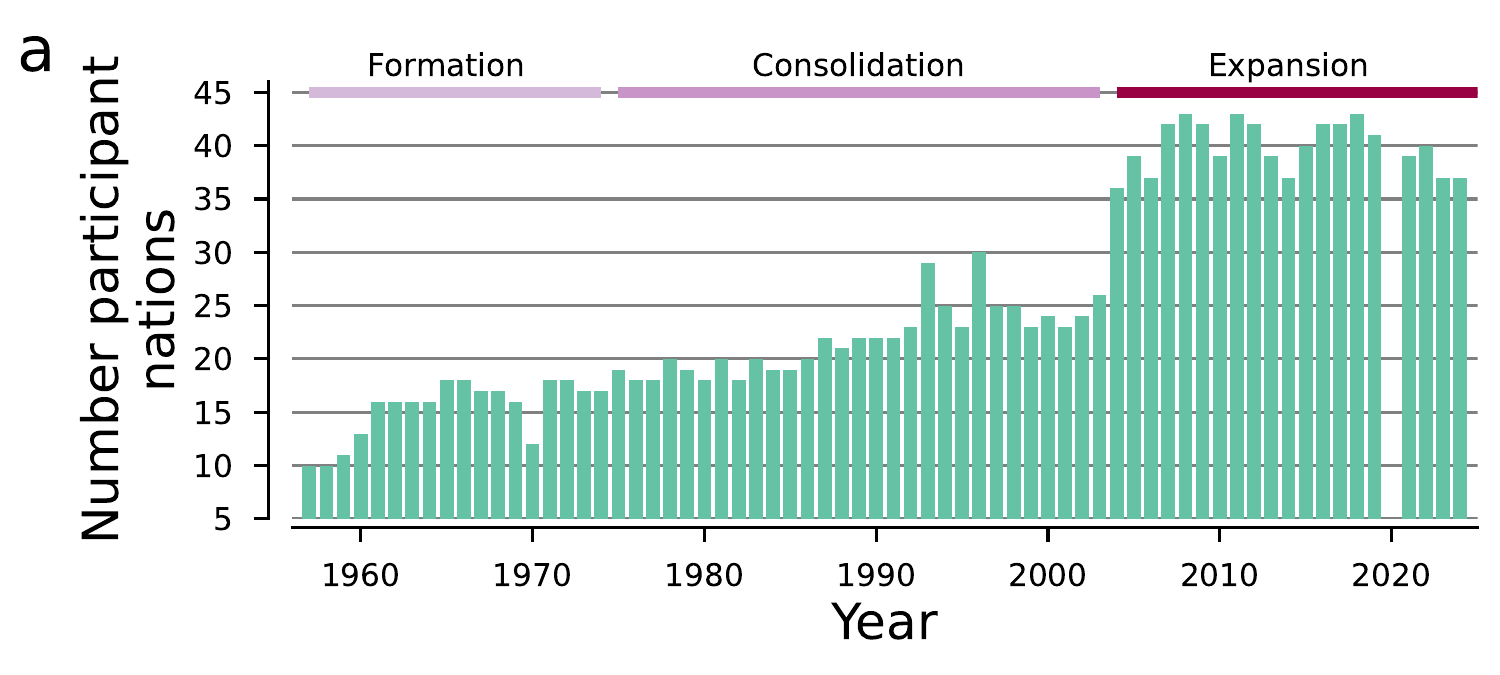} 
        \includegraphics[width=\linewidth]{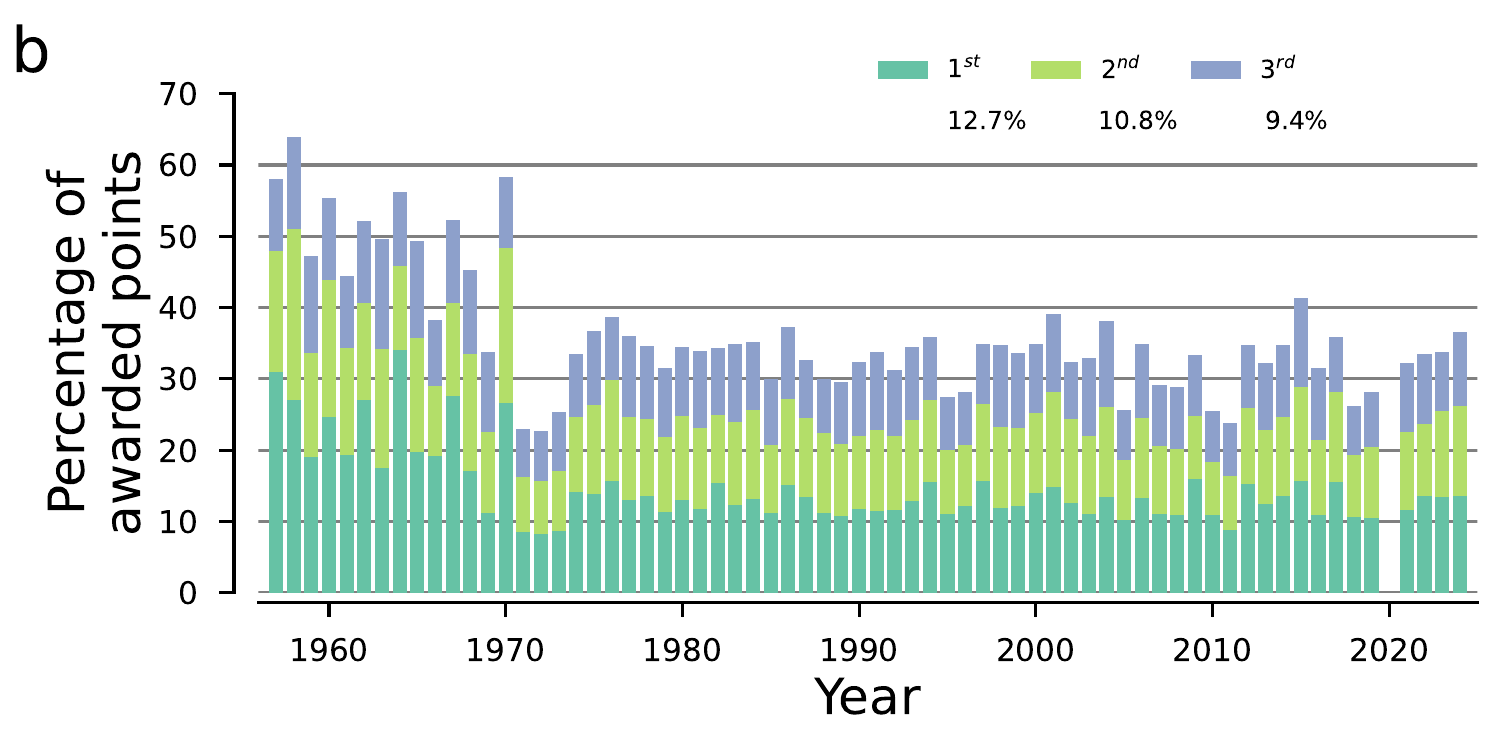} &  
        \includegraphics[width=\linewidth]{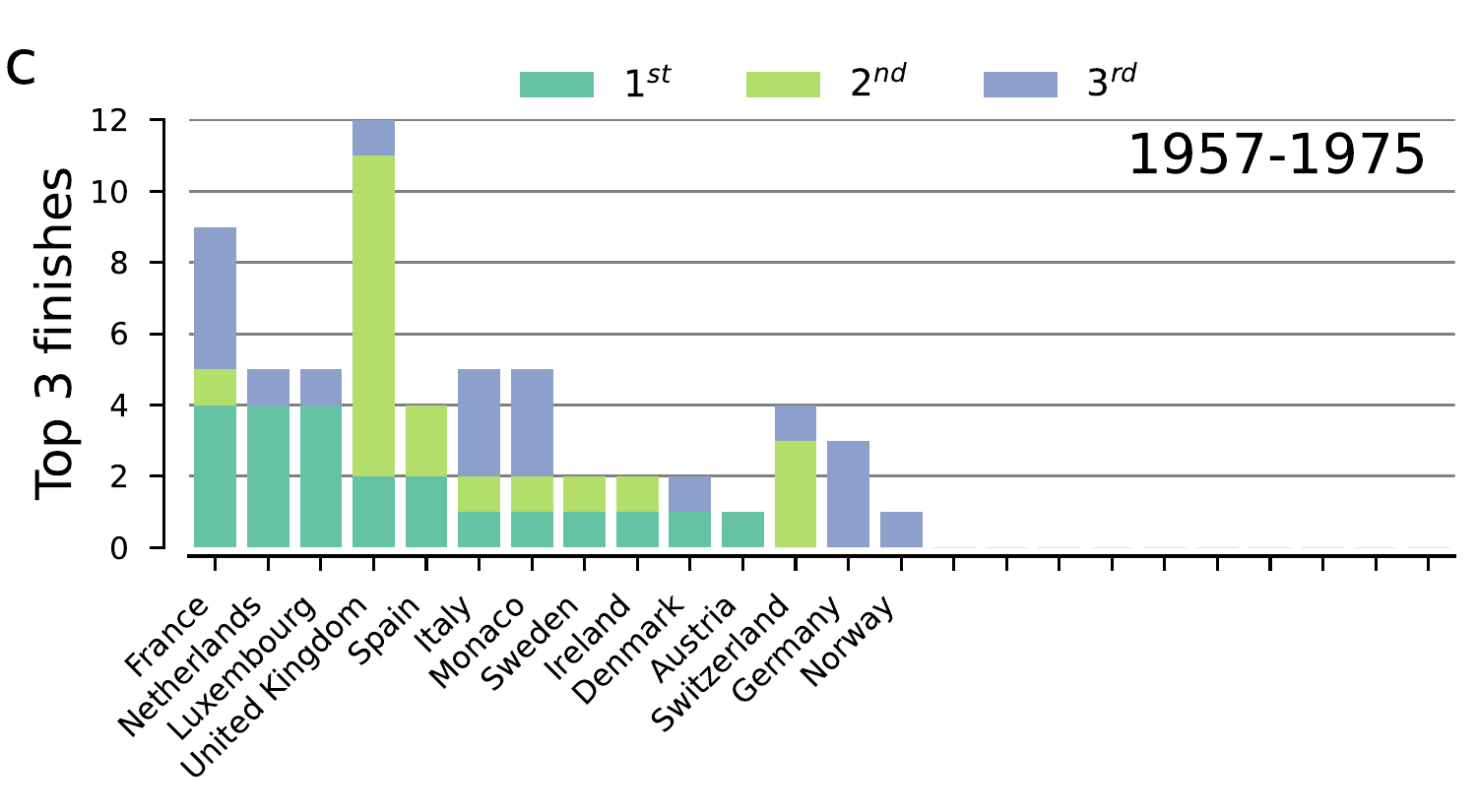}
        \includegraphics[width=\linewidth]{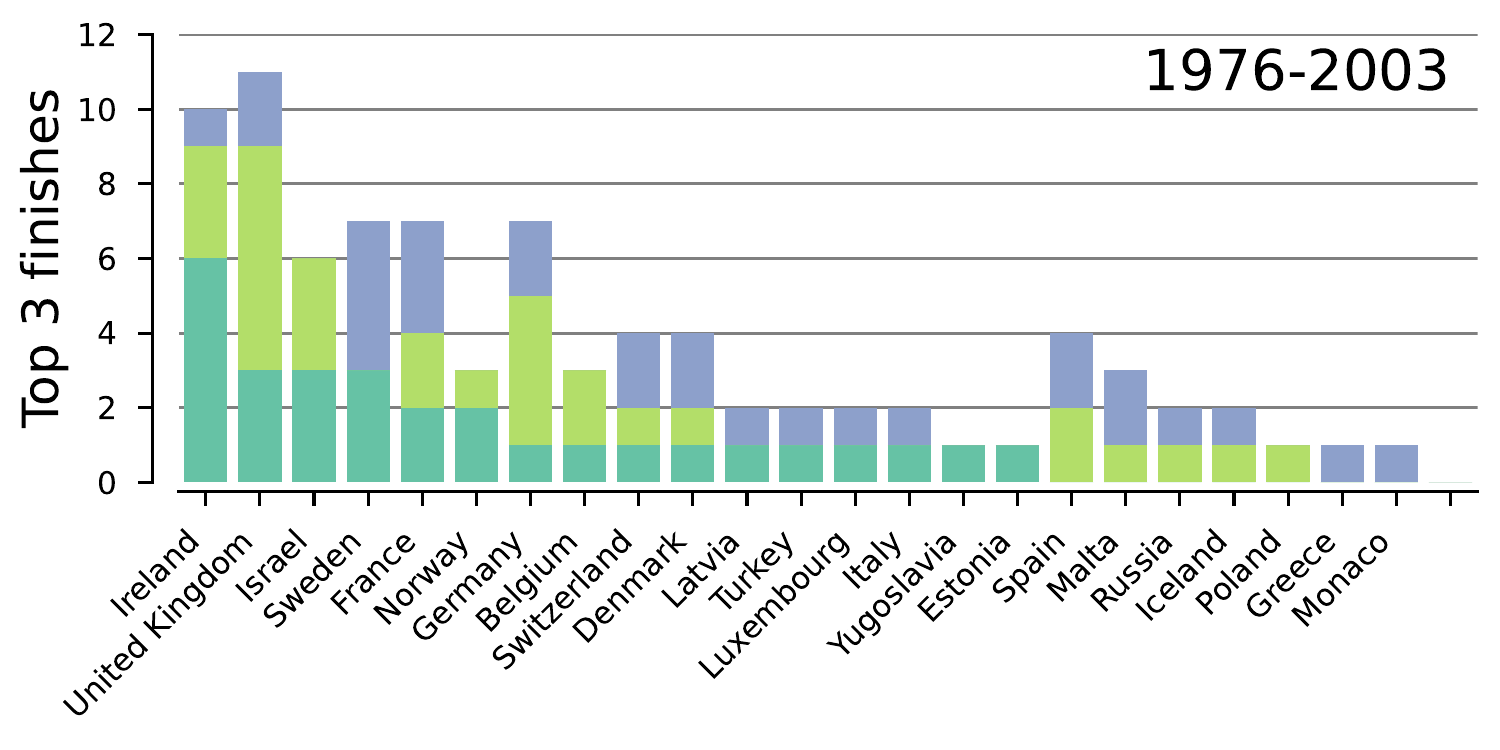}
        \includegraphics[width=\linewidth]{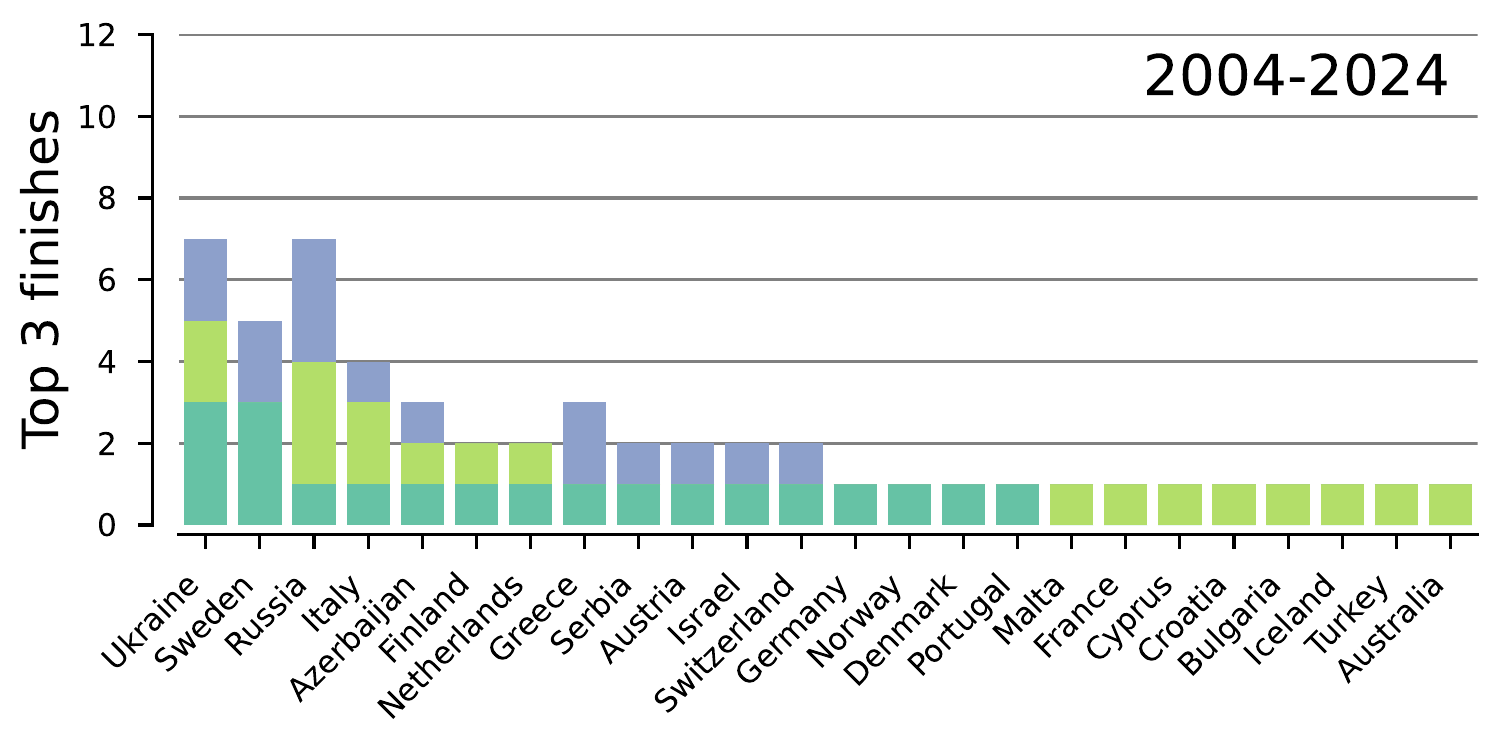} \\
    \end{tabular}
    \caption{\textbf{Over its history, Eurovision has experienced dramatic changes in the range of nations participating and voting, in the rules underlying the contest, and in its outcomes.  These changes can be interpreted response of the organizers to an increased homogeneity of songs competing.} 
    {\bf a,} The number of countries participating in Eurovision has increased steadily over time.  A major change, however, occurred in 2004 when there was an increase in the number of participants and nations allowed to vote by nearly 50\%.  This increase required the use of semifinal contests. We split this 68-year history into 3 periods that we denote "Formation", "Consolidation", and "Expansion".
    {\bf b,}  Changes in the voting process resulted in changes in the distribution of the votes received by participating nations (see Fig.~S\ref{fig:vote_distribution}). Importantly, it also affected the fraction of total votes received by the top 3 performing songs. It is visually apparent, nonetheless, that the system reached a stationary state after 1974. 
    {\bf c,} While the voting share of the winners remained stationary after 1974, the voting patterns did not. The top panel here shows the number of top three finishes for the countries participating in the 1957 to 1975 Contests.  It is visually apparent that some countries systematically do well while many others do not. Indeed, the top performing 4 countries took the majority of top 3 finishes. This concentration of winning decreases over time to almost parity in the Expansion period.  This evolution suggests that the advantage held by some countries initially was lost over time . 
    }
    \label{fig:eurovision}
\end{figure}

We start by extracting aggregated metrics for each Contest (Fig.~\ref{fig:eurovision}). We find that the number of nations participating in the Contest show three periods, which we denote as {\it Formation\/}, {\it Consolidation\/}, and {\it Expansion\/} (Fig.~\ref{fig:eurovision}).  The ``Formation'' period includes the Contests taking place prior to 1974. During this period, the number of participating nations increased from fewer than 10 to just over 15.  The voting was not yet standardized and changed very much from year to year and from participant to participant (see Fig.~\ref{fig:vote_distribution} for the impact of these changes).  The second period, ``Consolidation'', includes the Contests from 1974 to 2003.  While there is a steady growth in the number of participating nations from about 15 to about 25, the voting rules remain stable. This results also in a stable partitioning across Contests in terms of the fraction of all votes earned by each of the top three songs (Fig.~\ref{fig:eurovision}b). The third period, ``Expansion'', includes the Contest from 2004 onward.  Its start coincides with a very large increase in the number of participating nations and in the number of voting countries. Furthermore, a semi-final stage was introduced for most of the participating countries. Finally, from 2004 there has been an increase in the impact of televoting. These changes result in a greater degree of variability in the vote fraction earned by the top three songs (Fig.~\ref{fig:eurovision}b) and in a significant increase in the entropy of the distribution of votes (Fig.~\ref{fig:vote_distribution}).

The initial group of participating nations---Belgium, France, Germany, Italy, Luxembourg, Netherlands and Switzerland --- are located in close proximity, but differ in a number of cultural aspects, including language.  By 1975, the set of participating nations also included the Scandinavian countries, the Iberian countries, Ireland, Turkey, the U.K, and former Yugoslavia, thereby dramatically increasing the number of languages and cultures represented. One could surmise that such diversity might lead to the formation of voting blocks centered around language, religion, or geographic proximity. To test this hypothesis, we investigate whether there are expected or unexpected patterns in how a given nation awards its available votes.
%
\begin{figure}[b!]
    \centering
    \includegraphics[width=0.55\linewidth]{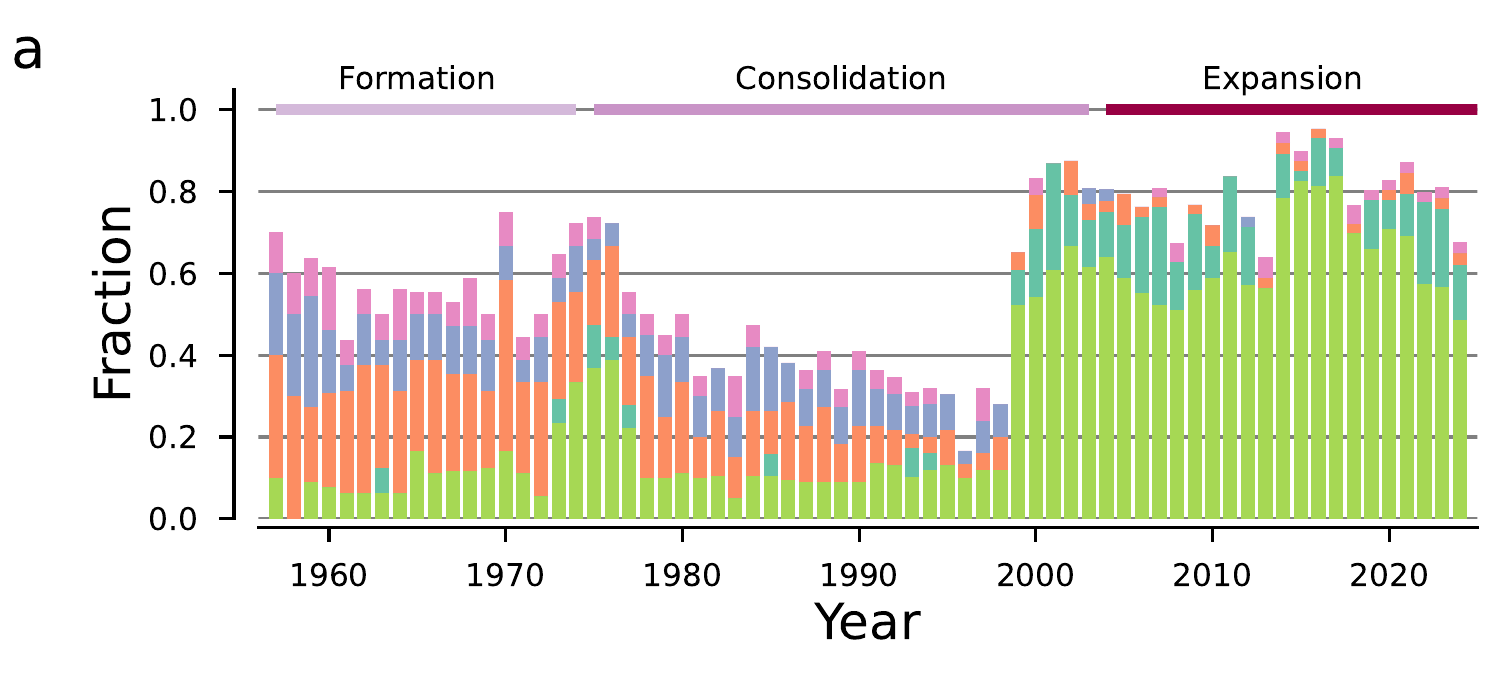} \\
    \includegraphics[width=0.55\linewidth]{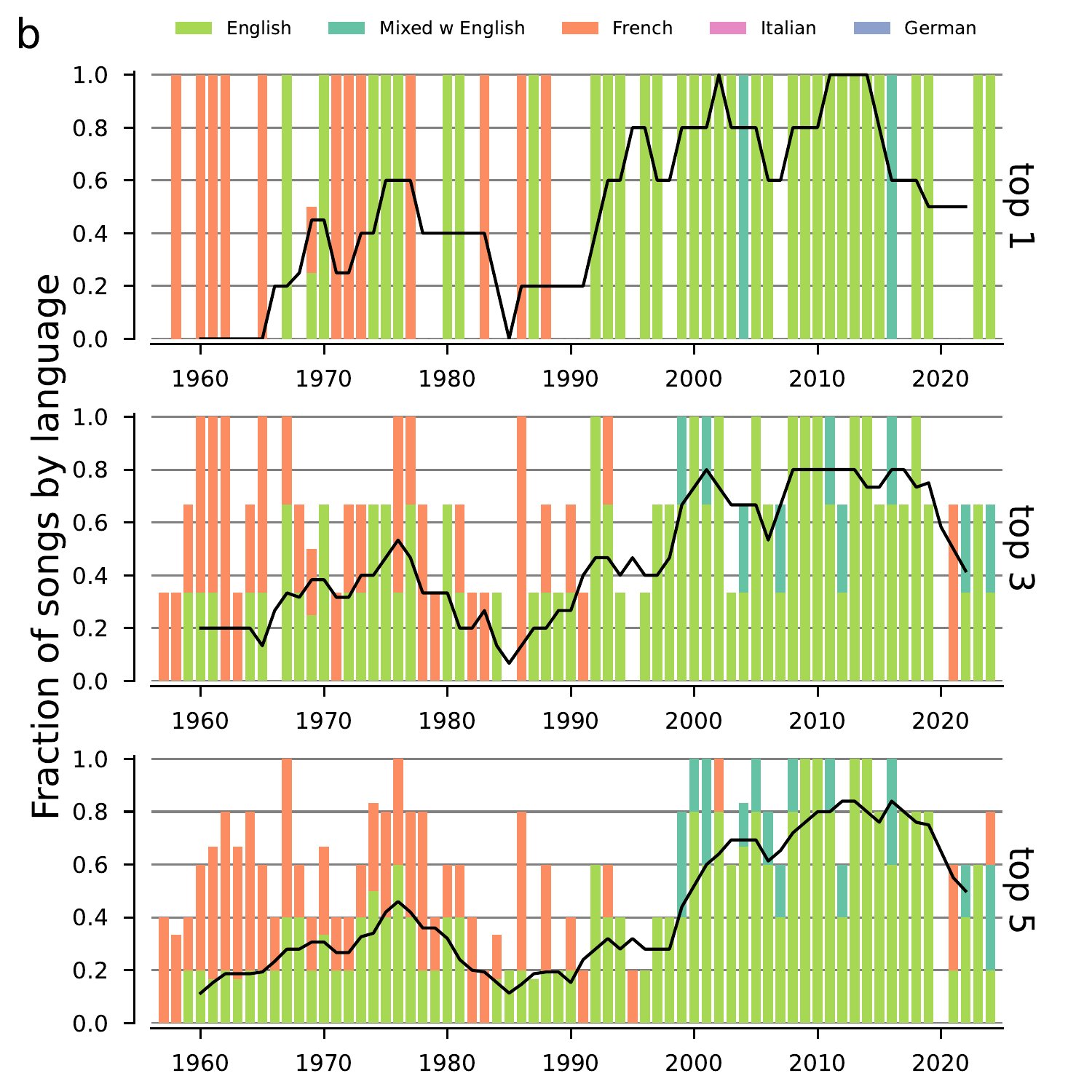}
    \caption{ \textbf{Despite the diversity of European languages, English becomes the \textit{de facto\/} language of Eurovision.}  
    {\bf a.} Fraction of songs using the languages of most of the initial competing countries. During the "formation" period, there is a great diversity in the language of the competing songs. This diversity increases through most of the "Consolidation" period. Suddenly, in 1998, there is a transition to a new state, where English becomes the dominant language. In this new state of affairs, over 70\% of songs are entirely in English or a mix of English and another language. 
    {\bf b.} When considering the languages of the top performing songs, the situation is quite different.  Initially, French is the dominant language but, already by the 1970s, nearly 40--60\% of the winning songs are in English (see full black line, which shows 5-year running average).  By the early 1990s, nearly 80\% of winning songs are in English. While the rate of increase is less pronounced for the top 3 and top 5, the data is consistent with the learning by participants that songs with English lyrics have an advantage. The increased dominance of English as the language of the top 3 songs is accompanied by a decrease of French, which nearly disappears as the language of choice among the top 5 songs after 1990. }
    \label{fig:language}
\end{figure}

Figure~\ref{fig:eurovision}c shows that the rate at which different countries place in the top 3 positions changes over time. In the Formation period, top placement in concentrated within 4 countries, France, Netherlands, Luxembourg, and the U.K. This concentration may be due to a number of factors, including voting rules.  However, during the Consolidation and Expansion stages, the voting rules become set, making an interpretation of the results easier.  Thus, it is still surprising that top 3 placement is rather more concentrated during the Consolidation period then during the Expansion period.  


\subsection*{Language}

If the population speaking a song language would  explain these voting consensuses, then it would be expected that bigger nations such as France, Germany and the U.K. should perform particularly well (Fig.~\ref{fig:english_speakers}).
In order to test this hypothesis, we calculate the fraction of songs with lyrics using a given language for each year (Fig.~\ref{fig:language}a). These data imply  two striking observations. First, there is a steady decline over time in the fraction of songs that have French or German lyrics. Second, in 1999 there is a sudden, step-like increase in the use of English lyrics. Unlike the rise in the use of English around the mid-1970s, this rise shows no sign of reversing.  So, we may ask, why did this step change happen?

To answer this question, we calculate how the fraction of top ranked songs that have English lyrics changes over time.  Figure~\ref{fig:language}b makes clear that --- whether we consider the winning song, the top three songs, or the top 5 songs --- by 1970 the top-rated songs are primarily using English. Moreover, while the best performing songs, prior to 1970, had frequently used French lyrics, French lyrics became a strong predictor of ``not top 5'' after 1990.

These results make it clear that, during the Consolidation period, participating countries learned that submitting songs with lyrics in English dramatically increased the competitiveness of the song. Strikingly, this learning, of a pattern that was already visible earlier, was adopted in mass only in 1999.  

Next, we investigate whether there are other winning song features that are being learned over time. Two features likely to be important in determining the popularity of a song are the characteristics of the lyrics, including complexity and theme, and the music attributes, including audio features and music genre.

\subsection*{Lyrics}

\begin{figure}[th!]
    \centering
    \includegraphics[width=0.55\linewidth]{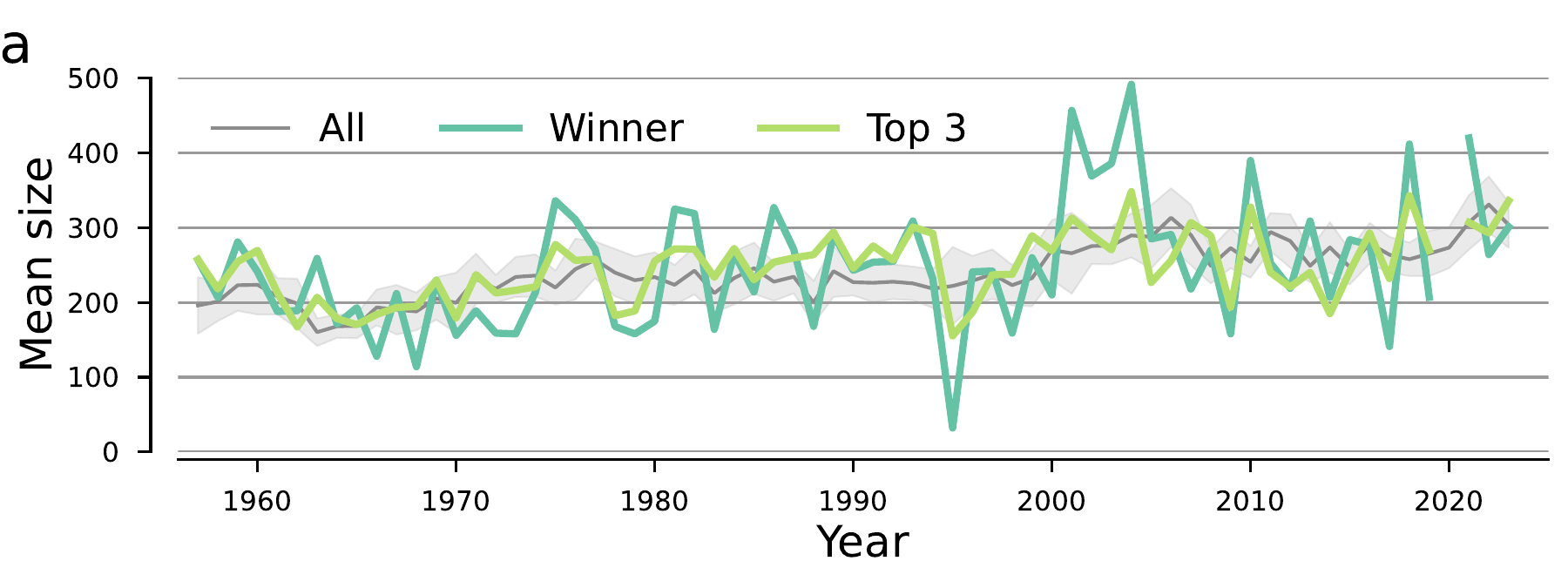} \\
    \includegraphics[width=.55\linewidth]{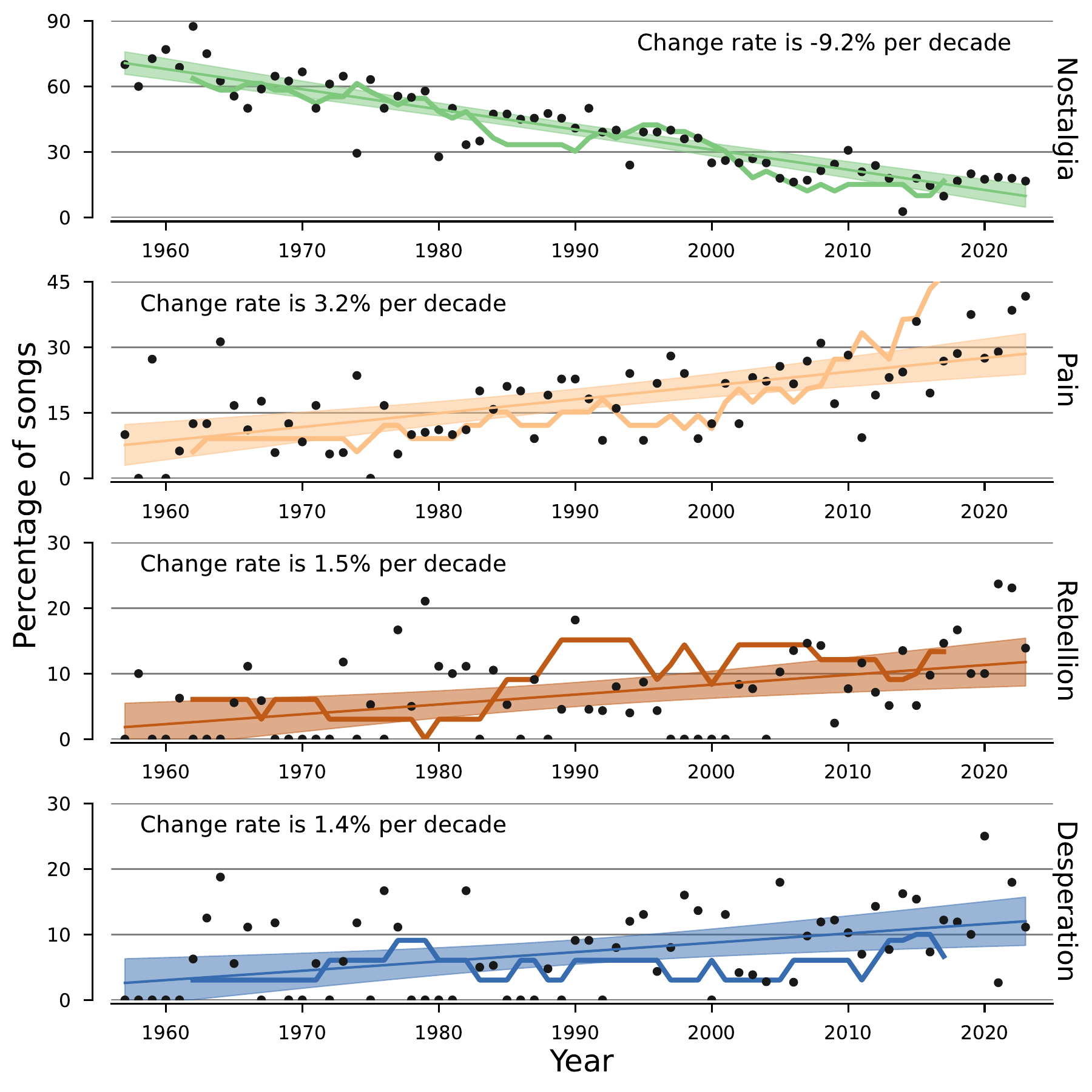}
    \caption{ {\bf Song lyrics also show sings of change over time consistent with participant learning.} 
    {\bf a.} Time evolution of mean lyrics size (as measured by number of words) for all songs, winning song, and Top 3 songs. For all songs, the shaded region sows the 95\% confidence interval for the estimate of the mean. As for song language, we find a step-like increase in lyrics size in 1999. Demonstrating the presence of a signal to be learned, we find that the lyrics size of Top 3 songs was already increasing during the period 1975--1998.  Interestingly, the winning songs, are characterized by extreme lyrics sizes, either very small or very large. As shown in Fig.~\ref{fig:lyric_sizes}, there is no significant dependence of lyrics size on country.
    {\bf b.} Time evolution of the percentage representation of the four most informative song themes for all songs and Top 3 songs. The black circles show the average over all competing songs for a given year, the line and shaded regions show 99\% confidence intervals for the linear fit, and the colored lines show running 10-year running average of theme representation among Top 3 songs. All four themes display strong and statistically significant temporal trends. While top performing songs follow the common trend for Nostalgia, they focused less on Pain between 1990 and 2000, but since 2010 are focusing more on Pain than the average of all competing songs. }
    \label{fig:lyrics}
\end{figure}

First, we focus on the characteristics of song lyrics. A straightforward measure of lyrics complexity is the size of the lyrics as measured by number of words. Figure~\ref{fig:lyrics}a shows the time evolution of the mean lyrics size. By 1975, once rules of the competition stabilize and some learning has had opportunity to occur, lyrics stabilize around a size of about 230 words. This range is valid across countries (Fig.~\ref{fig:lyric_sizes}). However, by 1999, there is a step-like 20\% increase in lyrics size.  As for song language, we find that songs placing in the Top 3 already have larger lyrics size by the late 1980s.  However, the lyrics size of the winning song shows much greater fluctuations. 

The themes present in song lyrics also display interesting patterns. The 11 most common themes occur in at least 5\% of all songs.  Moreover, their co-occurrence is not random.  Aspiration and Inspiration appear more frequently together than one would expect from chance alone, as do Pain, Breakup and Loss (Fig.~\ref{fig:topic_cooccurrences}).

Four song themes --- Nostalgia, Pain, Rebellion, and Desperation --- display significant temporal trends. Surprisingly, an initially strong focus on Nostalgia has been in steep decline.  This may be related to the increasing distance from the horrors of World War II and improving economic conditions. In contrast to Nostalgia's steep decline, Pain, Rebellion and Desperation have been increasing significantly in representation. Intriguingly, Confusion and Escapism show nearly a step change, stretching from 1970 to 1980, to higher levels of representation, whereas Loss shows an opposite change over the same period to lower levels of representation (Figs.~\ref{fig:topic_trends} and \ref{fig:topic_trends2}). This could be a response to all the crises of the 1970s.  

These findings suggest that the songs competing in Eurovision are picking up on emotional trends permeating culture. But a question is whether top performing songs merely ride these trends or whether they contribute to their emergence. The data is particularly interesting for Pain and Desperation (see also Fig.~\ref{fig:topic_tops}). Between 1990 and 2000, top performing songs refer less to Pain than the average competing song. Since 2000, however, the representation of Pain among top performing songs has been increasing steadily and exceeds the average annual level since 2010. It may not be a coincidence that this is the time post the Great Recession.

In contrast to Pain, Desperation has had lower representation among top performing songs than average songs.  This may be due to the emotional weight of a feeling such as desperation, which could turn voters off. 

\begin{figure}[b!]
    \centering
    \includegraphics[width=0.55\linewidth]{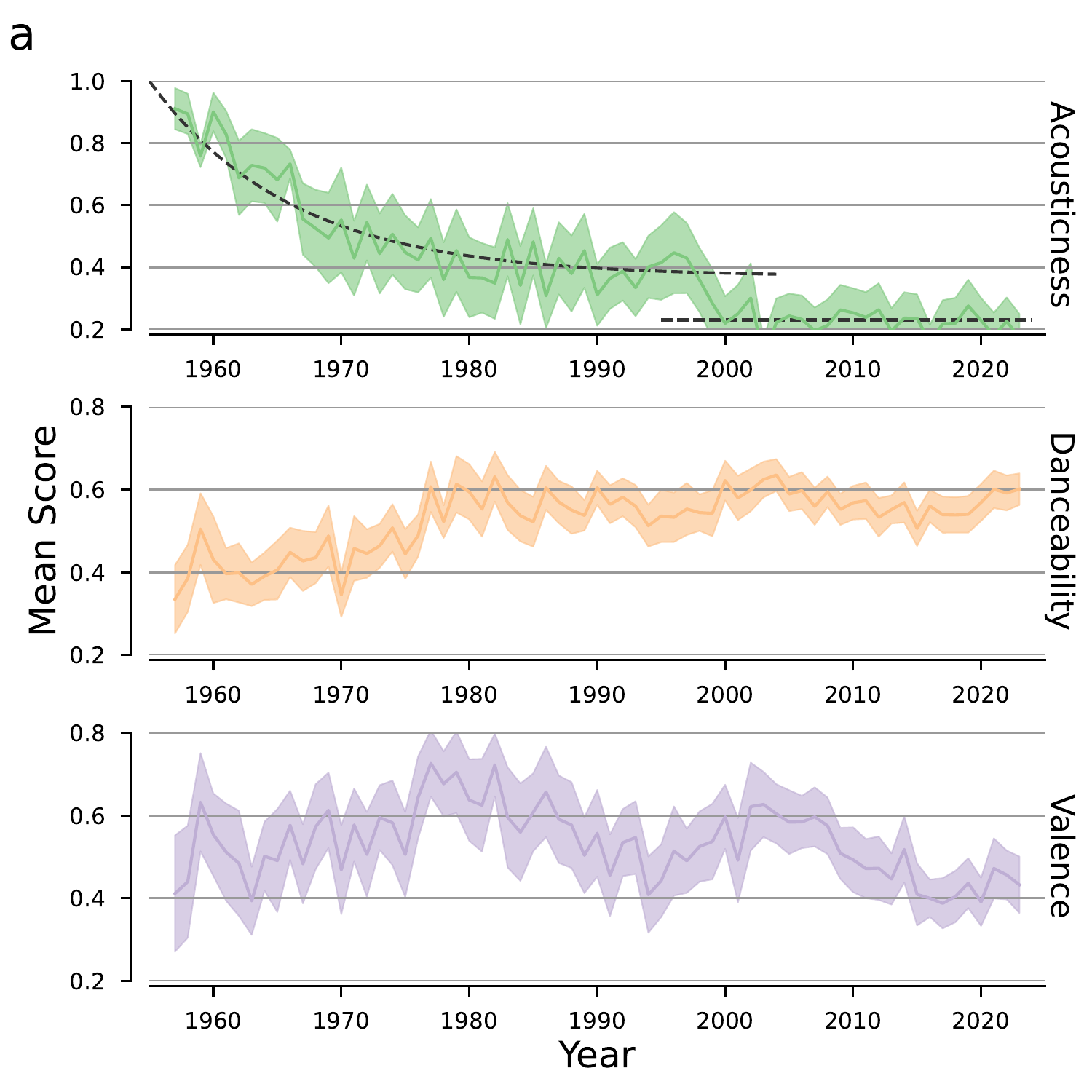} \\
    \includegraphics[width=0.55\linewidth]{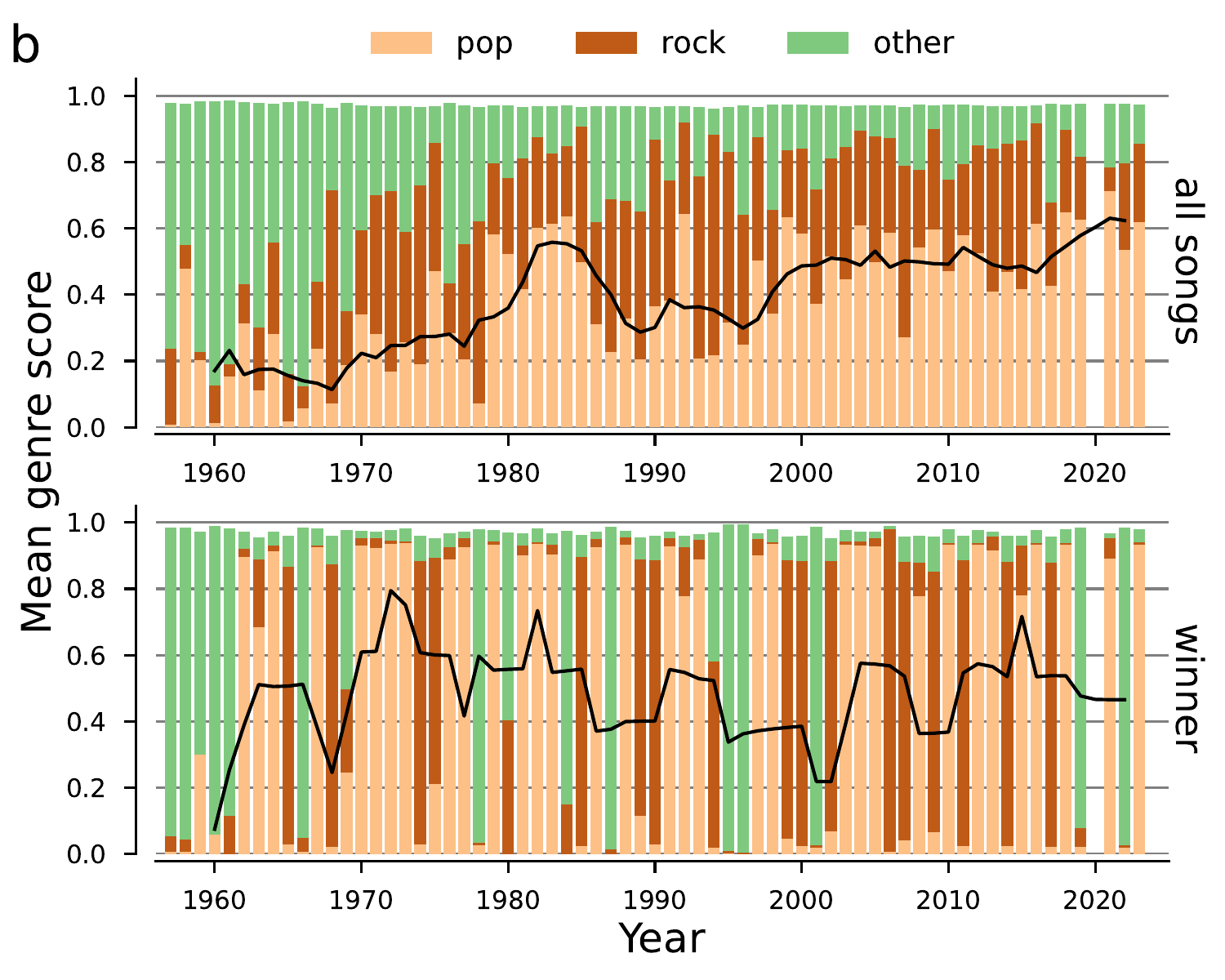}
    \caption{ \textbf{Despite the diversity of music styles and traditions in Europe, pop has become the dominant music genres for participating songs, but not for winning songs.}  
    {\bf a.} Annual mean audio feature scores over time of competing songs. The solid lines show the mean value of the attribute for the songs competing in a given year, while the red shaded areas display the $95\%$ confidence interval for the mean. The gray lines in the rightmost plot show fits to the functional forms discussed in the text.
    {\bf b.} Mean annual genre score of all competing songs and winning songs. The mean pop score of competing songs has been increasing steadily over the existence of the competition. Intriguingly, the genre of the winning songs shows much different trends (see black line, which shows the running average for 5-year time windows).  Since about 1970, half of the winning songs have been pop songs.  This pattern is also present when considering the top 3 or top 5 songs (Fig.~\ref{fig:music_genre_top3-5}) but occurs at a slower rate. These results suggest that, unlike language, where English is the standard, there is more opportunity for ``bucking the trend" in song genre. } 
    \label{fig:music}
\end{figure}

\subsection*{Music}

Now, we address the songs' audio features and genres.  Figure~\ref{fig:spotify_features_all} shows the distribution and cross-correlations for six audio features identified by the Spotify API~\cite{spotify_developer}.  It is visually apparent that several are strongly correlated, thus, we focus on just three of them that we believe provide the greatest insight: Acousticness, Danceability, and Valence. When considering their time evolution, some patterns clearly emerge (Fig.~\ref{fig:music}a). Acousticness, which was dominating earlier on, has been decaying for the entire existence of the Contest. At first, the decline was smooth and exponential in nature, but then, it happened in a step-like way. The early primacy of Acousticness has since been taken over by Danceability. Valence and Danceability follow a similar oscillating pattern. They both increase until they reach a peak in the late 1970s.  Then, they both decrease until the mid 1990s and  increase towards an early 2000s peak. That peak is again followed by a decrease until the mid 2010s, and an increase since.

The time evolution of audio features in competing songs is mirrored in the change in music genre. Figure~\ref{fig:music}b shows the mean genre score for Pop and Rock for all songs and for the winning song. It is visually apparent that Pop songs have come to dominate the Contest, although not as much as the English language, and not in such an abrupt manner. As for language and lyric characteristics, there is a clear signal regarding the characteristics of the top placed songs that allow all participants to learn. This is visible by the fact that the fraction of pop songs among the top performing echelon occurs earlier (Fig.~\ref{fig:music_genre_top3-5}). Interestingly, the situation is not as unequal as when considering language with regard to the genre of the winning song. Pop songs nowadays tend to make up about half of top performing songs.  

\begin{figure}[b!]
    \centering
    \includegraphics[width=0.5\linewidth]{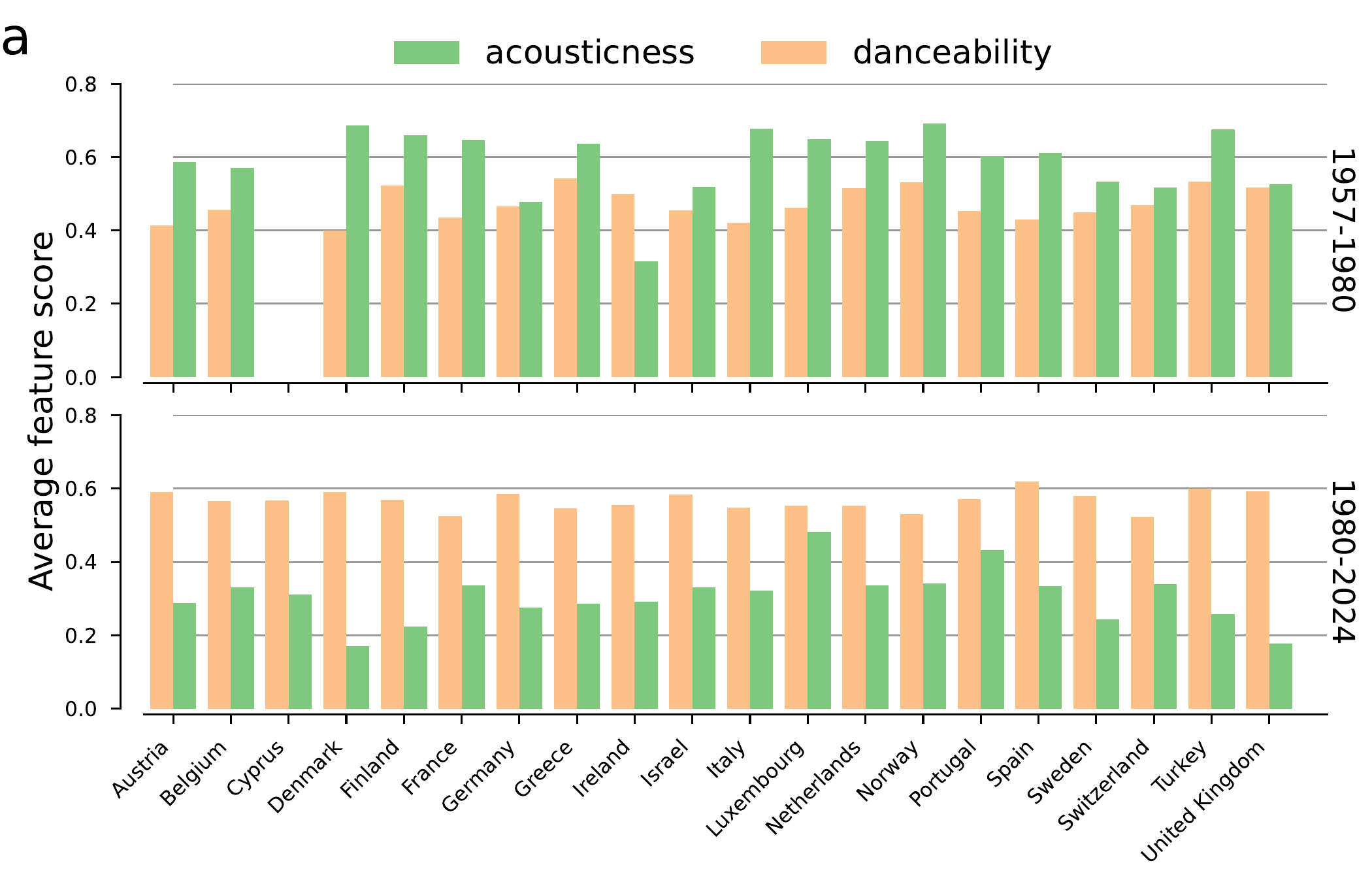 }
    \includegraphics[width=0.4\linewidth]{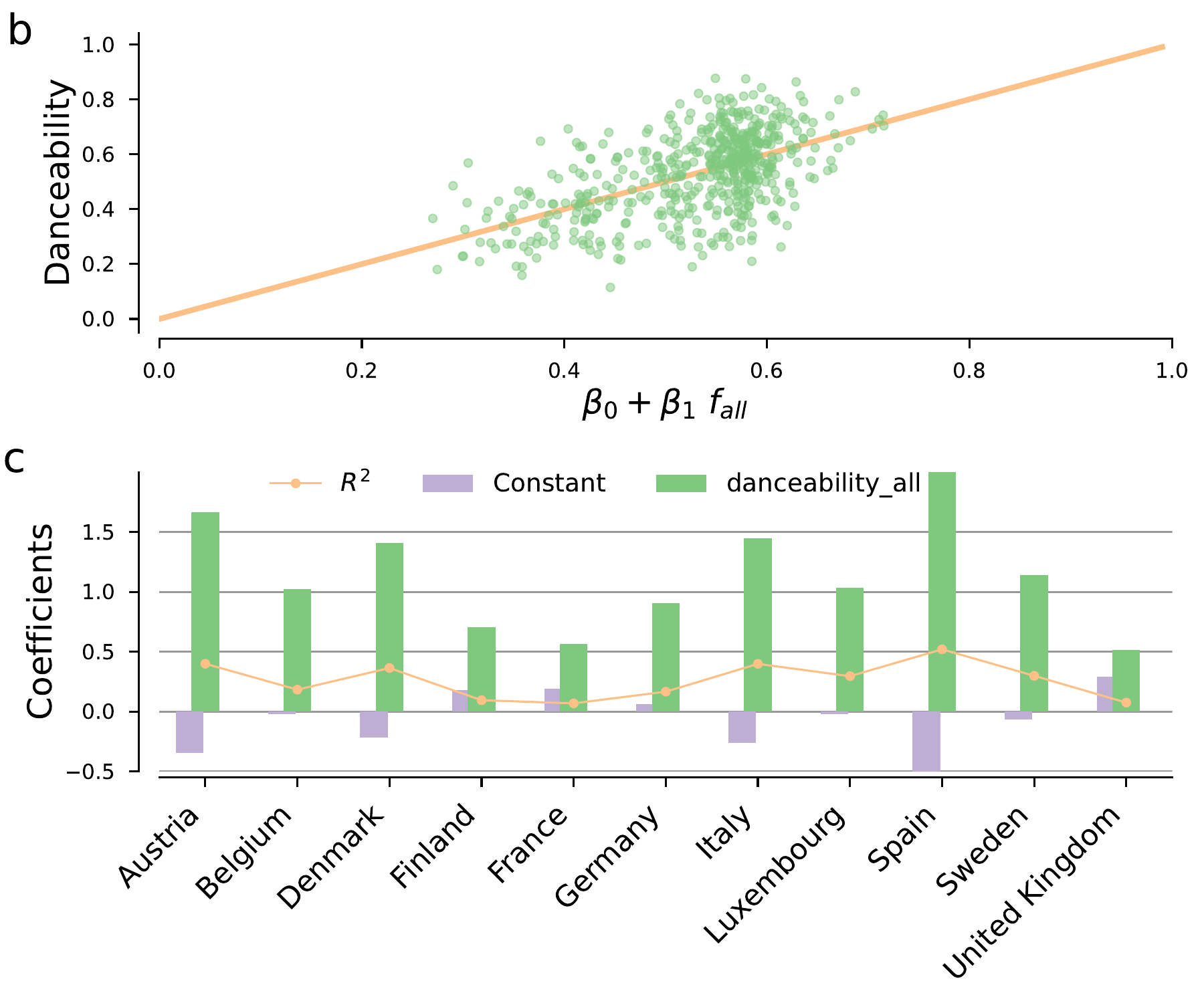 }
    \includegraphics[width=0.5\linewidth]{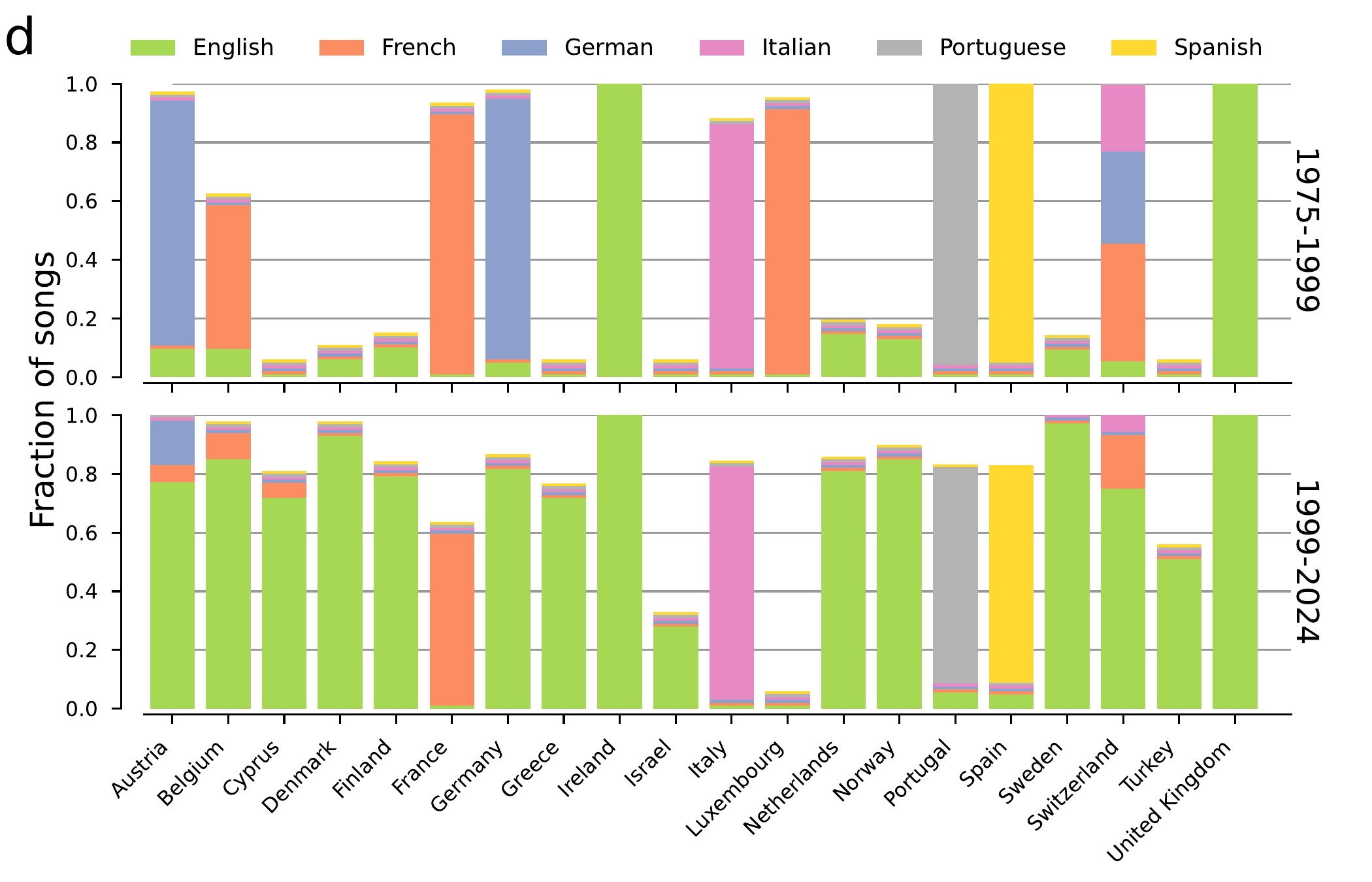}
    \includegraphics[width=0.40\linewidth]{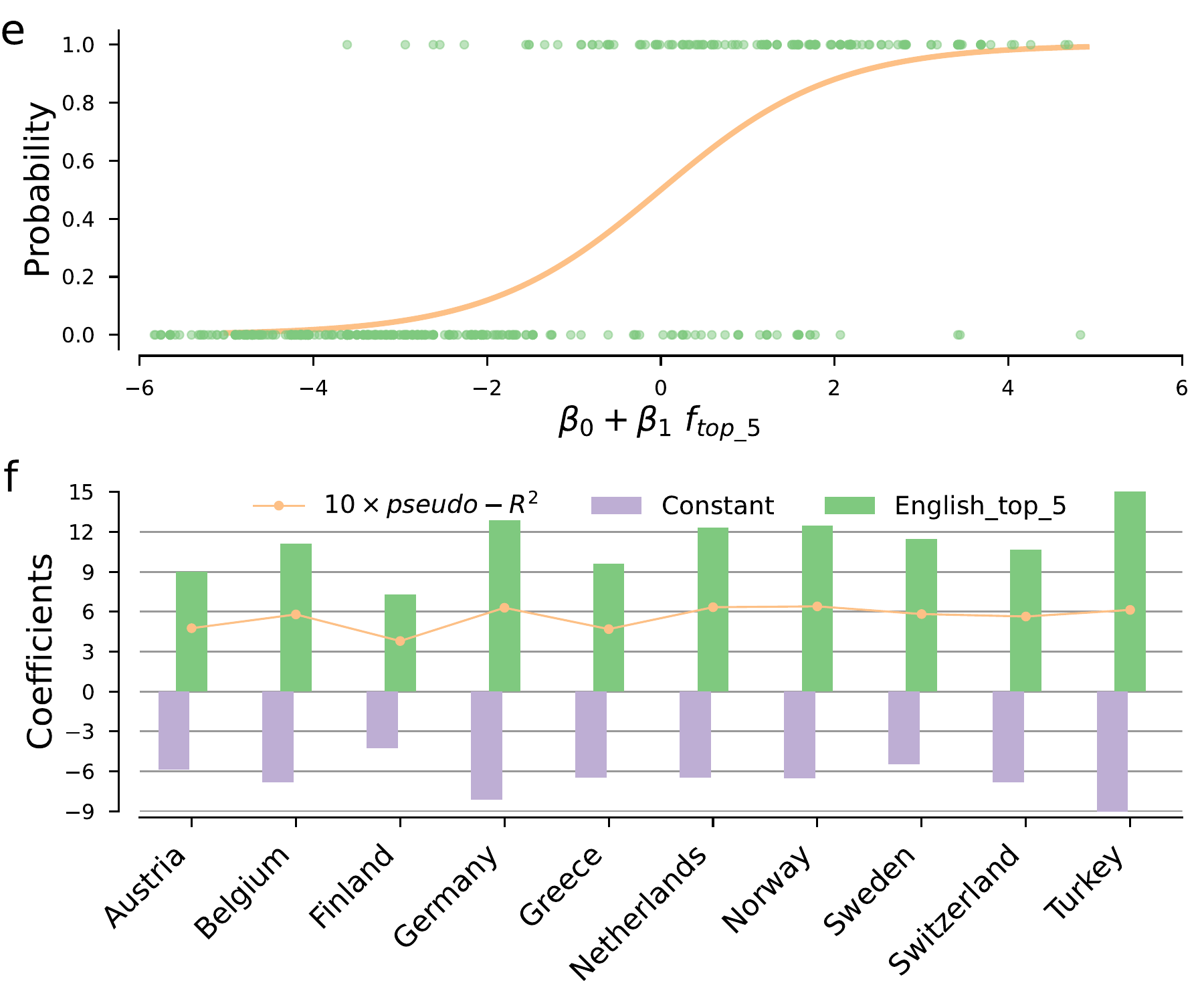 }
    \caption{ \textbf{Participant learning of winning strategies.} 
    {\bf a.}  Average Acousticness and Danceability scores of songs submitted to Eurovision by different countries across two time periods. It is visually apparent that, prior to 1980, competing songs had high Acousticness, but after 1980, competing songs --- across all countries --- had high Danceability. We find similar results for lyric sizes (Fig.~\ref{fig:lyric_sizes}).
    {\bf b.} Linear regression of danceability of a country's song in year $y$ against the average danceability of {\it all\/} other countries in the previous 3 years.
    {\bf b.} Regression coefficients for the 11 countries for which the linear model is statistically significant and $R^2$ of the linear fit. 
    {\bf d.} Fraction of songs with lyrics in a given language submitted to Eurovision nations across two times periods. Prior to 1999, most nations submitted songs in one of their national languages (see Switzerland pre-1999). Post-1999, most countries submitted songs with English lyrics. The exceptions were France, Italy, Portugal and Spain, which were mostly submitting songs in their national languages.
    {\bf e.} Logistic regression of a country's song having English lyrics in year $y$ to the average fraction of songs placing in the {\it top 5\/} with English lyrics in the previous 3 years.
    {\bf f.} Regression coefficients for the 10 countries for which the logistic model is statistically significant and  pseudo-$R^2$ of the fit. 
    }
    \label{fig:national_goals}
\end{figure}

\subsection*{Country specific learning}

Up to now, we have focused on system-wide learning. We found that competing countries have learned from each other about lyrics language, size and topic, and about audio characteristics and music genre.  What we have not investigated is whether this learning has occurred equally across countries or whether some countries have eschewed some of those lessons. Figure~\ref{fig:national_goals}a compares the usages of Acousticness and Danceability by country for the pre- and post-1980 periods. As expected, we find adoption of Danceability across the board. As shown in Fig.~\ref{fig:lyric_sizes}, we further find a similar adoption for lyric sizes across the board.

To determine whether these changes regarding music characteristics are due to learning, we fit a linear model to the danceability score of a country's song in year $y$ to the average danceability scores of different subsets of competing songs --- top 1, top 3, top 5, or all --- in the previous $n = 1, 3, 5$ years (Fig.~\ref{fig:national_goals}b).  We find similar statistically significant models for 11 of 20 countries (Fig.~\ref{fig:national_goals}c), revealing that these countries adjust the danceability of their songs in response to average danceability of {\it all\/} other countries in the past 3 years. We find similar results for acousticness (Fig.~\ref{fig:regression_acousticness}).

Figure~\ref{fig:national_goals}d shows language utilization by country for the period pre and post system-wide transition to English. Surprisingly, it is visually apparent that for the pre-1999 Contests, countries mostly submitted songs in their official languages. Post-1999, the lyrics of the songs submitted by most countries switched to English.  To determine whether these changes regarding song language are due to learning, we fit a logistic model to the probability that a country's song's lyrics are in English in year $y$ to the average fraction of Songs with English lyrics different subsets of competing songs --- top 1, top 3, top 5, or all --- in the previous $n = 1, 3, 5$ years (Fig.~\ref{fig:national_goals}e). We find similar statistically significant models for 10 of 18 countries (Fig.~\ref{fig:national_goals}f), revealing that these countries select the language of their songs in response to the fraction of songs in English of the {\it top 5\/} songs in the past 3 years.

The exceptions to English hegemony are France, Italy, Portugal, and Spain. The other four countries for which the logistic model fails to reproduced the data either have too strict a switch (Cyprus and Denmark), submit songs with a mixture of languages (Israel), or stopped competing after 1993 (Luxembourg).  It is unlikely that the first four countries did not learn about the advantage provided by the use of English, prompting the question of why they eschewed them.  Indeed, the likely reason for why France, Portugal, or Spain stick with their own languages is that they are spoken by large numbers of people around the world.  Thus, paying a cost for promoting their own languages is a rational choice in a geopolitical context that extends beyond winning at Eurovision.\cite{yair2022march}

Finally, it is interesting to consider the extent to which learning, and the consequent cultural homogenization, is embraced by different countries.  If one considers its population size, language utilization in Europe, or economic might, Germany would appear to be a natural case for setting trends instead of following them. Instead, we find that France, Spain and Italy --- but especially Ireland and Portugal --- are the countries shunning peer pressure. It is striking that some of these countries have shown a more independent streak on some of the most controversial political matters facing Europe.

\section*{Discussion}

Our study provides insights into how learning and adaptation occurs at multiple levels in a cultural market. From the organizers, who aim to maximize television audiences by making the Contest's outcome unpredictable and exciting, to the participating countries, who want to maximize the chances that their representatives' songs win the Contest while attending to other geopolitical necessities, to the voters, who may want to be fair in the ratings but are influenced by cultural and linguistic baggage.  All these participants have left traces in the Contests' records that enable us to determine what they have collectively learned and how they have decided to use those lessons.

Organizers tweaked with voting rules during the Formation period of the Eurovision, fostering a more predictable process, but less predictable outcomes. These changes led to a stable Consolidation period, which has eventually pushed organizers to expand voting options and voting participants.  The recency of those changes does not yet allow us to provide final insights.  In particular, the impact of televoting and the possibility of its manipulation with ``bot farms" 
is not without precedent.  Such manipulation has been attempted in the context of political elections.\cite{berghel2018malice}
While the GDPR recently provides some level of protection against this, the desire for greater degrees of soft power is likely to keep up the risk of such manipulation, particularly in times of information wars and propaganda.

On the participant side, we have revealed the generalized adoption of strategies that increase song competitiveness: use of English language, increased danceability, increased use of Pop styles, more diverse lyrics, and the focus on themes that better capture the cultural  zeitgeist. These results prompt the question of whether one could combine data analytics with generative AI to create songs that would ideally resonate with the diverse audience of Eurovision.  Whether that will be possible or not, one has to ask whether it would be desirable.  


\subsection*{Acknowledgments}
L.A.N.A. thanks the project ‘CoCi: Co-Evolving City Life’, which received funding from the European Research Council (ERC) under the European Union’s Horizon 2020 research and innovation programme under grant agreement no. 833168 for partial financial support of his research stay at ETH Zurich. 
A.T.E.C. acknowledges partial financial support by the European Union—Horizon 2020 Program under the scheme “INFRAIA-01-2018-2019—Integrating Activities for Advanced Communities”, Grant Agreement n.871042, “SoBigData++: European Integrated Infrastructure for Social Mining and Big Data Analytics” (http://www.sobigdata.eu).
D.H. has enjoyed coordinating and contributing to research in the CoCi and SoBigData++ projects mentioned above.

\newpage
\newpage
\clearpage

\bibliography{references}

\newpage
\newpage
\clearpage

\pagebreak
\newpage
\setcounter{figure}{0}
\makeatletter 
\renewcommand{\thefigure}{S\@arabic\c@figure}
\renewcommand{\thetable}{S\@arabic\c@table}

\newpage

\section*{Supplementary Information}

\begin{figure}[ht!]
    \centering
    \includegraphics[width=0.7\linewidth]{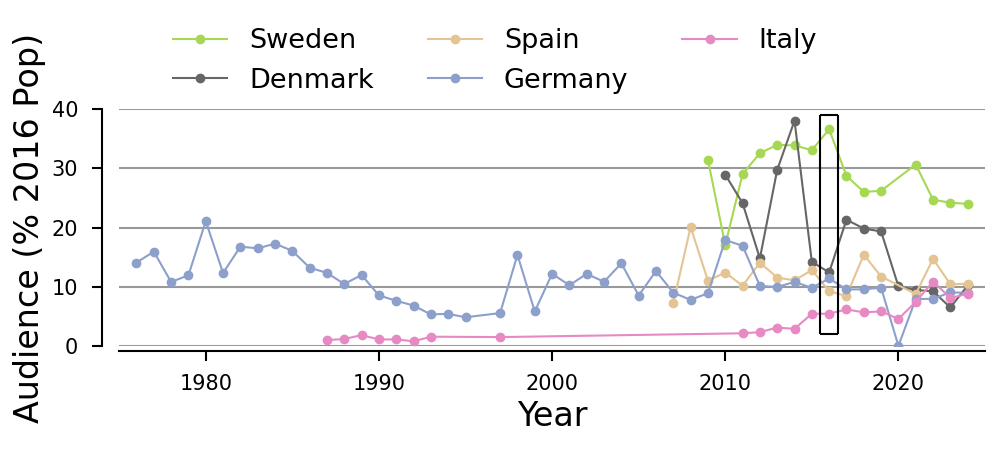}
    \includegraphics[width=0.7\linewidth]{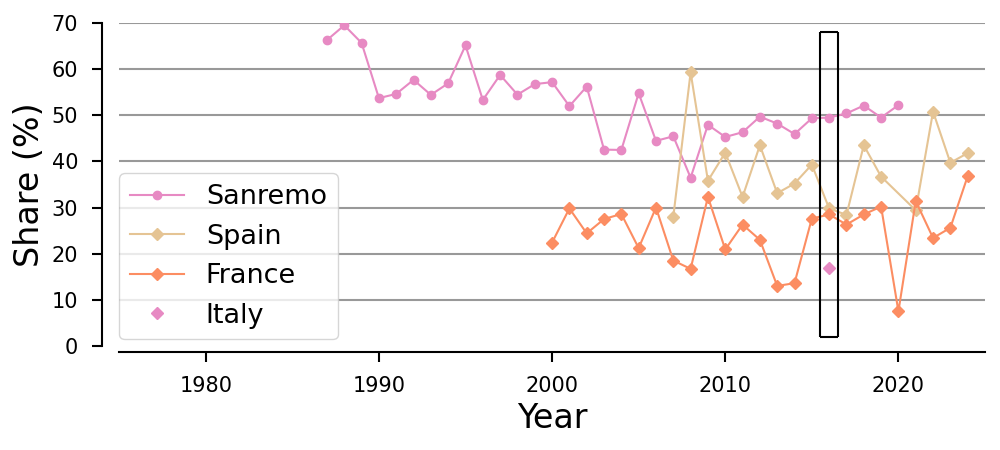}
    \caption{ \textbf{Eurovision's popularity, as measured by size of television audiences, has changed dramatically over time and across countries. }  We highlight 2016, the year for which we can make cross country comparisons. 
    {\bf a.} Televisions audiences for Eurovision at some of the participating countries as a percentage of the 2016 country population.  German audiences decreased dramatically after 1985, but grew in response to good performances by the German representative. Audiences from other countries seems to be converging to approximately 10\% of the population. But this convergence has been the result of decreases of popularity in some countries such as Denmark and increases in popularity in others countries (such as Italy). 
    {\bf b.} Eurovision television audience share for Spain and France compared to Sanremo Festival television audience share in Italy.  It is notable how much more popular the Sanremo Festival is in Italy than Eurovision is in France, Italy or Spain.}
    \label{fig:esc_audiences}
\end{figure}

\begin{figure}[ht!]
    \centering
    \includegraphics[width=0.9\linewidth]{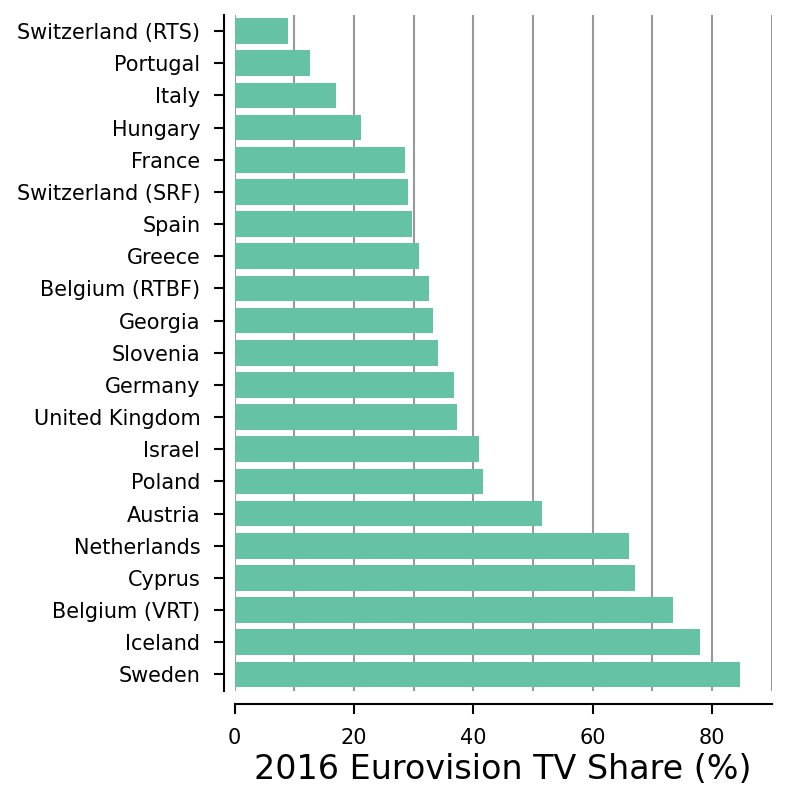}
    \caption{ \textbf{Television viewers share of 2016 Eurovision's final show across 18 European countries.} The viewers share of the Big Five, France, Germany, Italy, Spain, and the U.K., ranges from less than 20\% (Italy) to nearly 40\% (U.K.). }
    \label{fig:esc_2016_share}
\end{figure}

\begin{figure}[ht!]
    \centering
    \includegraphics[width=0.7\linewidth]{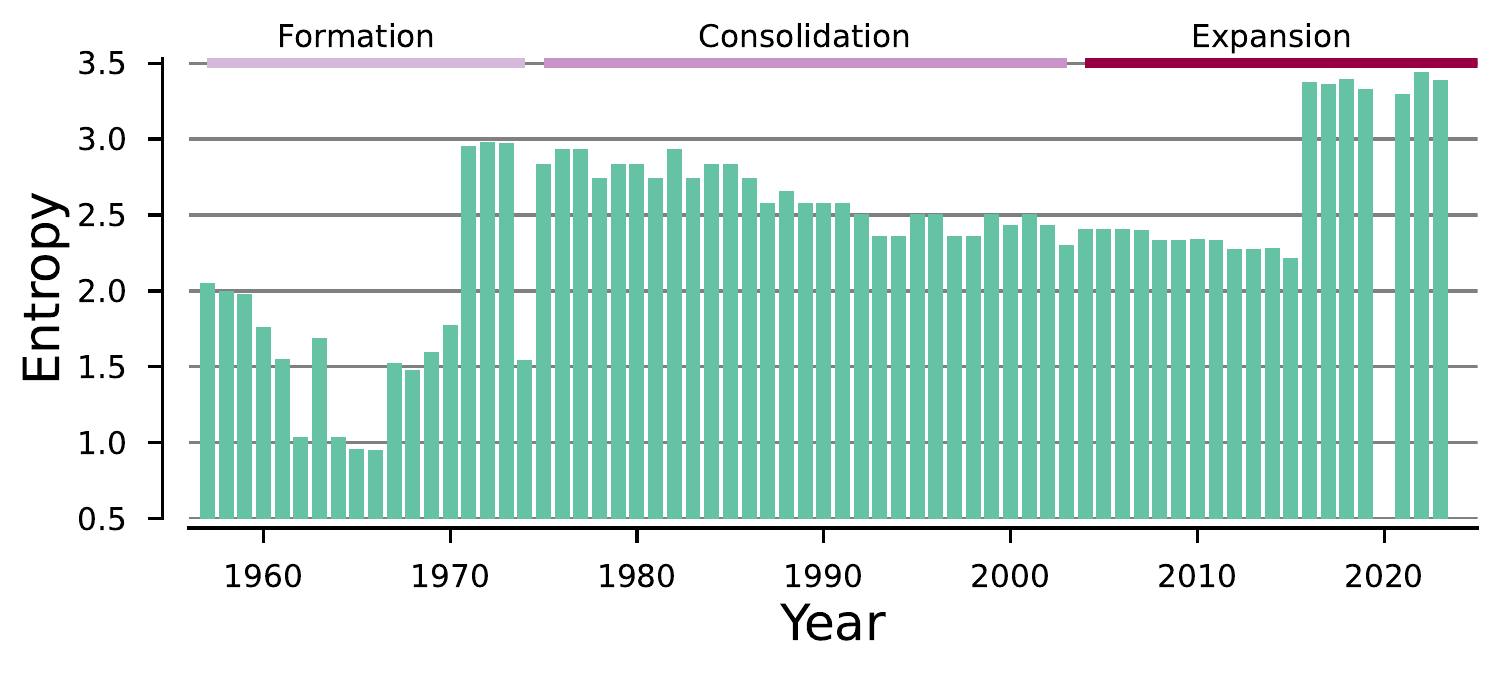}
    \caption{ \textbf{
    The distribution of votes changes as different voting rules are implemented and the number of participants increases.}  We calculate the entropy of the distribution of votes given 
    as a way to measure the heterogeneity of the distribution. A higher value indicates that a larger number of participants gets a significant share of votes.}
    \label{fig:vote_distribution}
\end{figure}

\begin{figure}[th!]
    \centering
    \includegraphics[width=\linewidth]{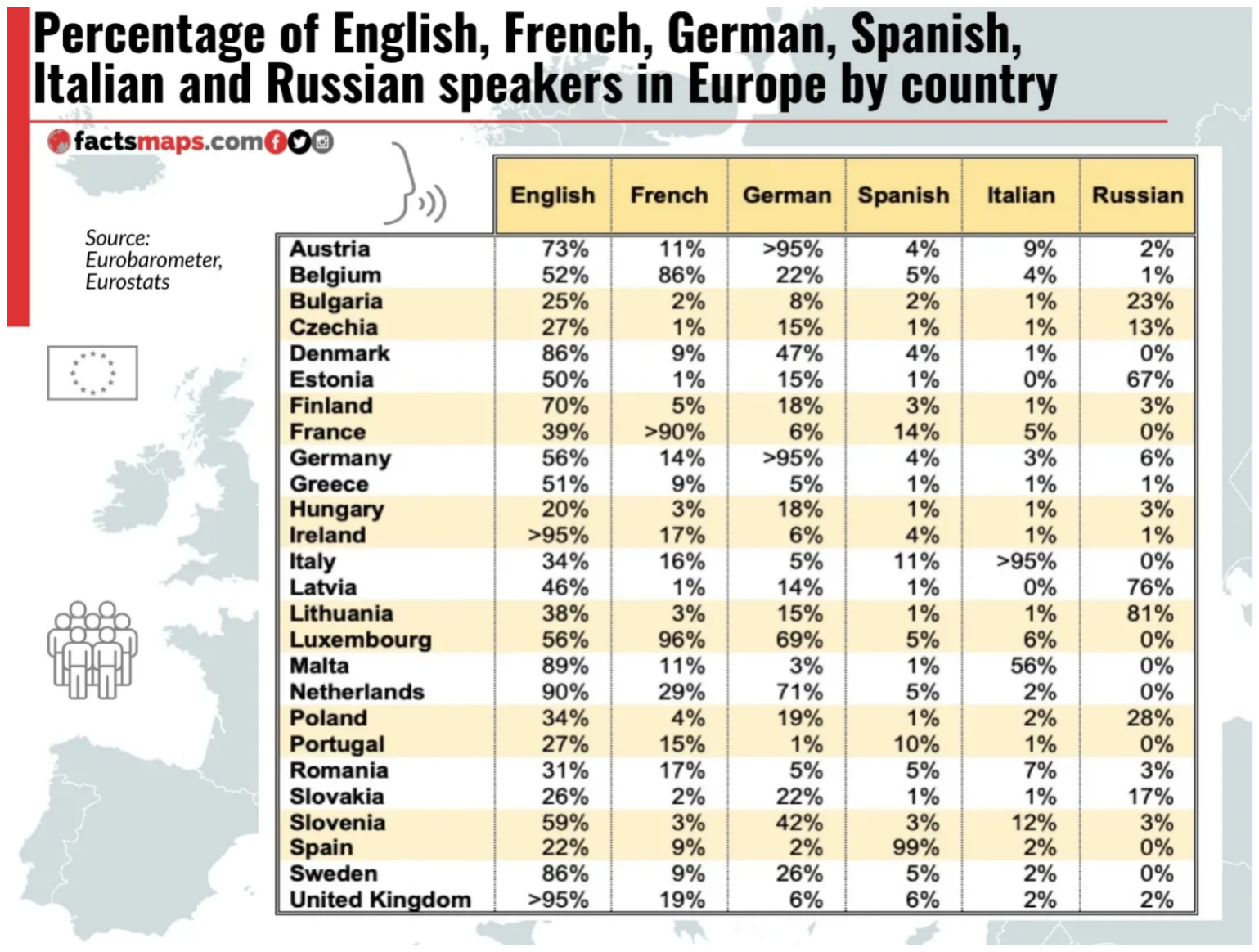}
    \caption{ {\bf Percentage of English, French, German, Spanish, Italian and Russian speakers in Europe by country.} After \href{https://factsmaps.com/percentage-of-english-french-german-spanish-italian-russian-speakers-in-europe/}{factsmaps.com}.  Notice how English speakers is a dominant second language in nearly all countries.}
    \label{fig:english_speakers}
\end{figure}

\begin{figure}[th!]
    \centering
    \includegraphics[width=0.6\linewidth]{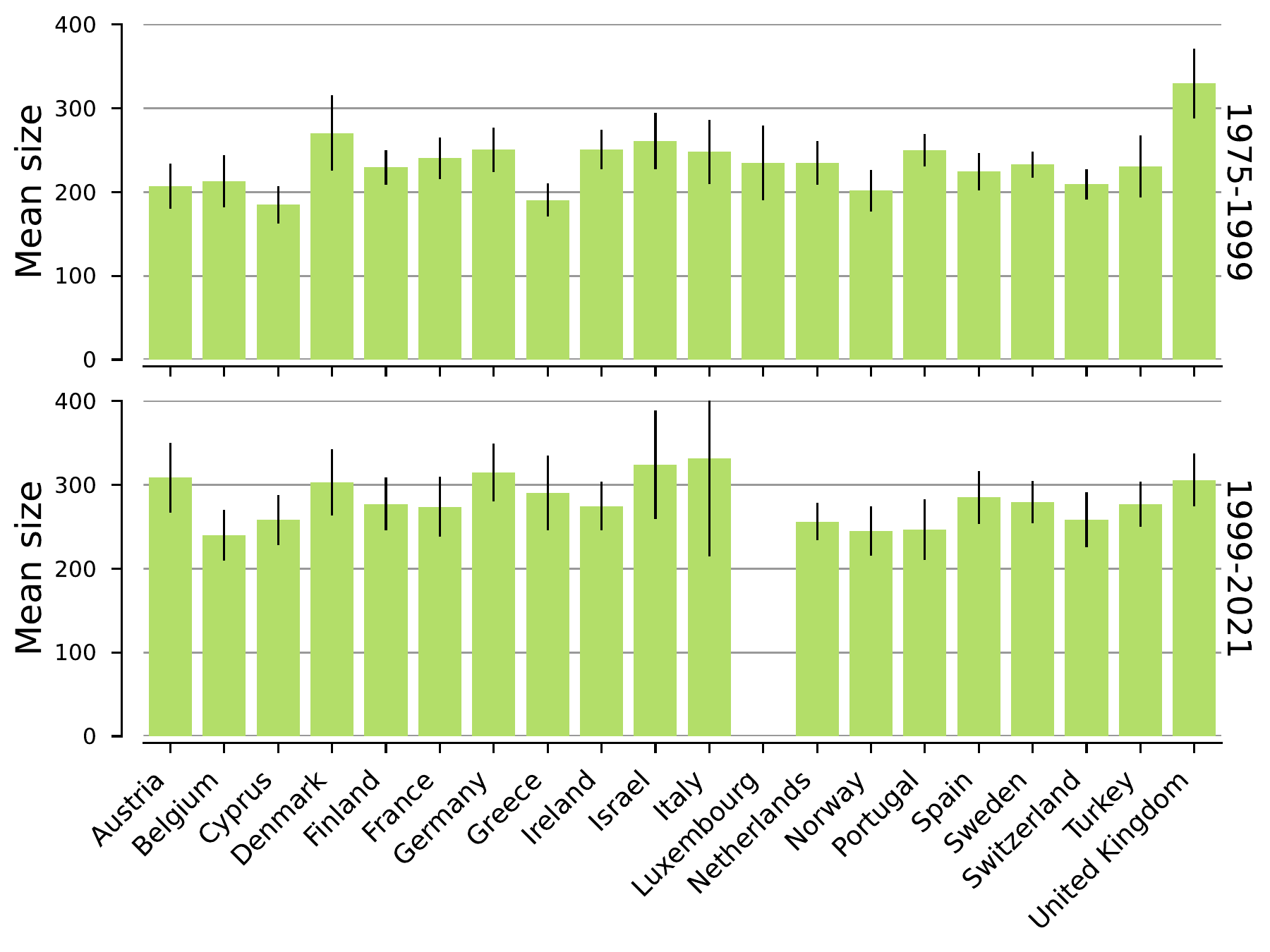}
    \caption{ {\bf Dependence of mean lyrics size on country for the pre- (1975--1998) and post-increase periods (post-1998).} It is visually apparent that, while there is no significant dependence of lyrics size on country, there is a significant dependence on time.}
    \label{fig:lyric_sizes}
\end{figure}

\begin{figure}[th!]
    \centering
    \includegraphics[width=0.9\linewidth]{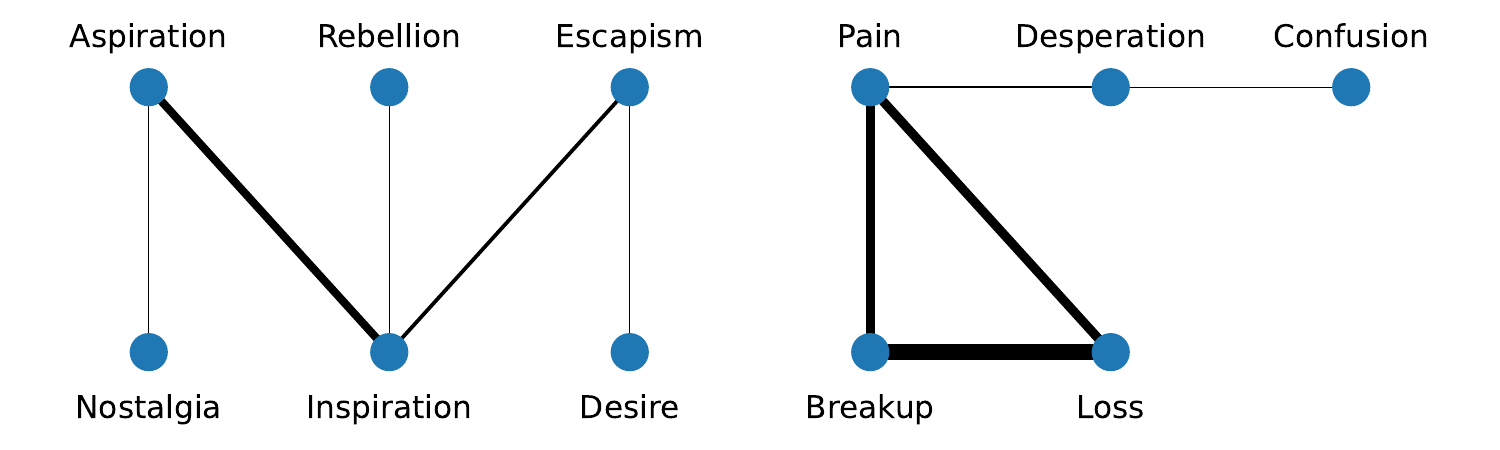}
    \caption{ \textbf{Co-occurrence of themes within songs.} We calculate the excess co-occurrences with regard to the null model of themes co-occurring at random.  Reassuringly, we find that Loss, Breakup and Loss co-occurs more frequently than would expect by chance. Similarly for Aspiration and Inspiration.} 
    \label{fig:topic_cooccurrences}
\end{figure}

\begin{figure}[th!]
    \centering
    \includegraphics[width=\linewidth]{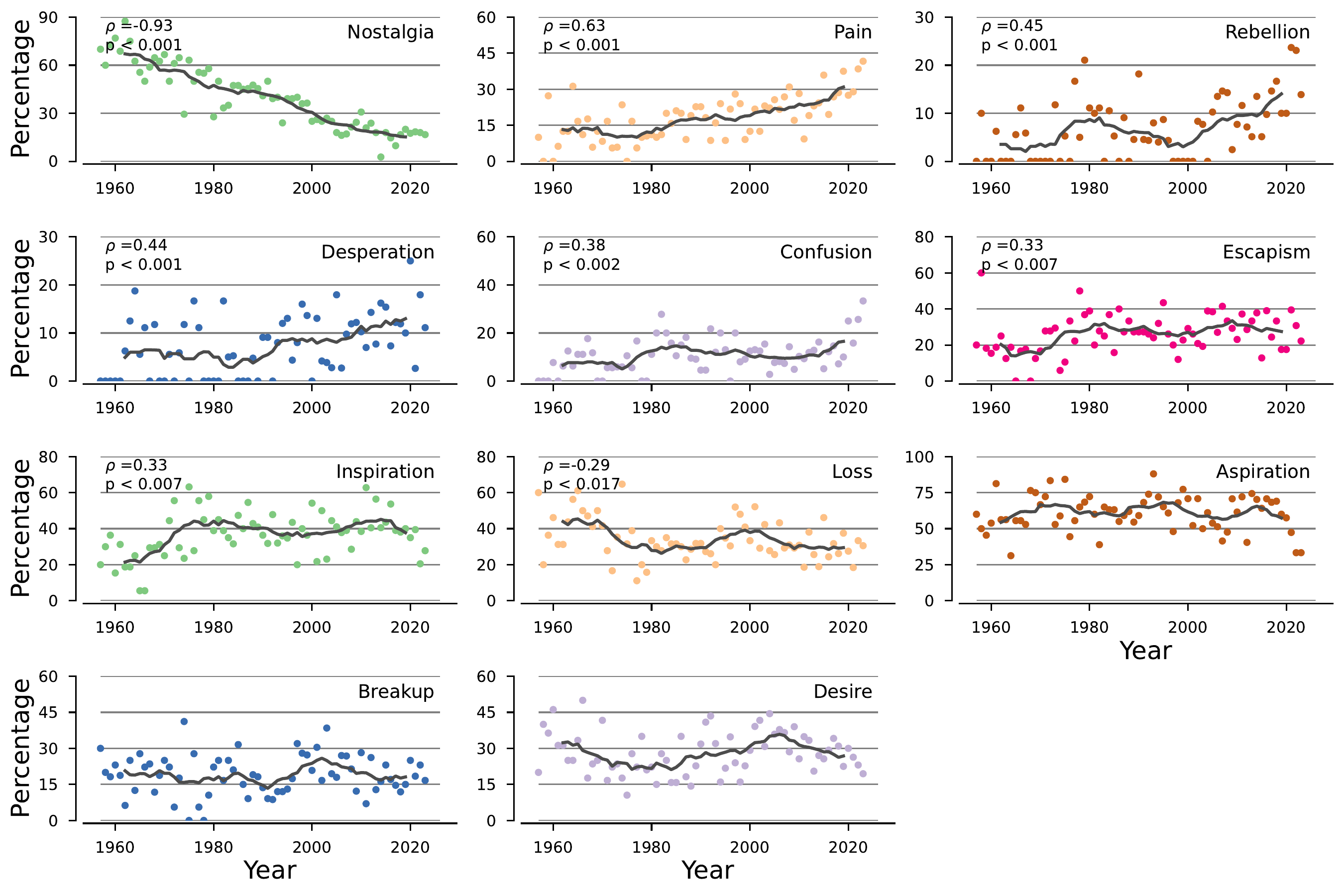}
    \caption{ \textbf{Temporal trends in theme usage by competing songs in the period 1957 to 2024. } We use GPT~4o to assign up to three themes to the lyrics of competing songs. The plots show the fraction of songs assigned individual songs annually.  We calculate the Spearman's rank correlation of this quantity with time, and its p-value, for the 11 themes broadly represented.  We display their these values for the cases where the $p$-value of the estimated correlation coefficient is smaller than 0.05. }
    \label{fig:topic_trends}
\end{figure}

\begin{figure}[th!]
    \centering
    \includegraphics[width=\linewidth]{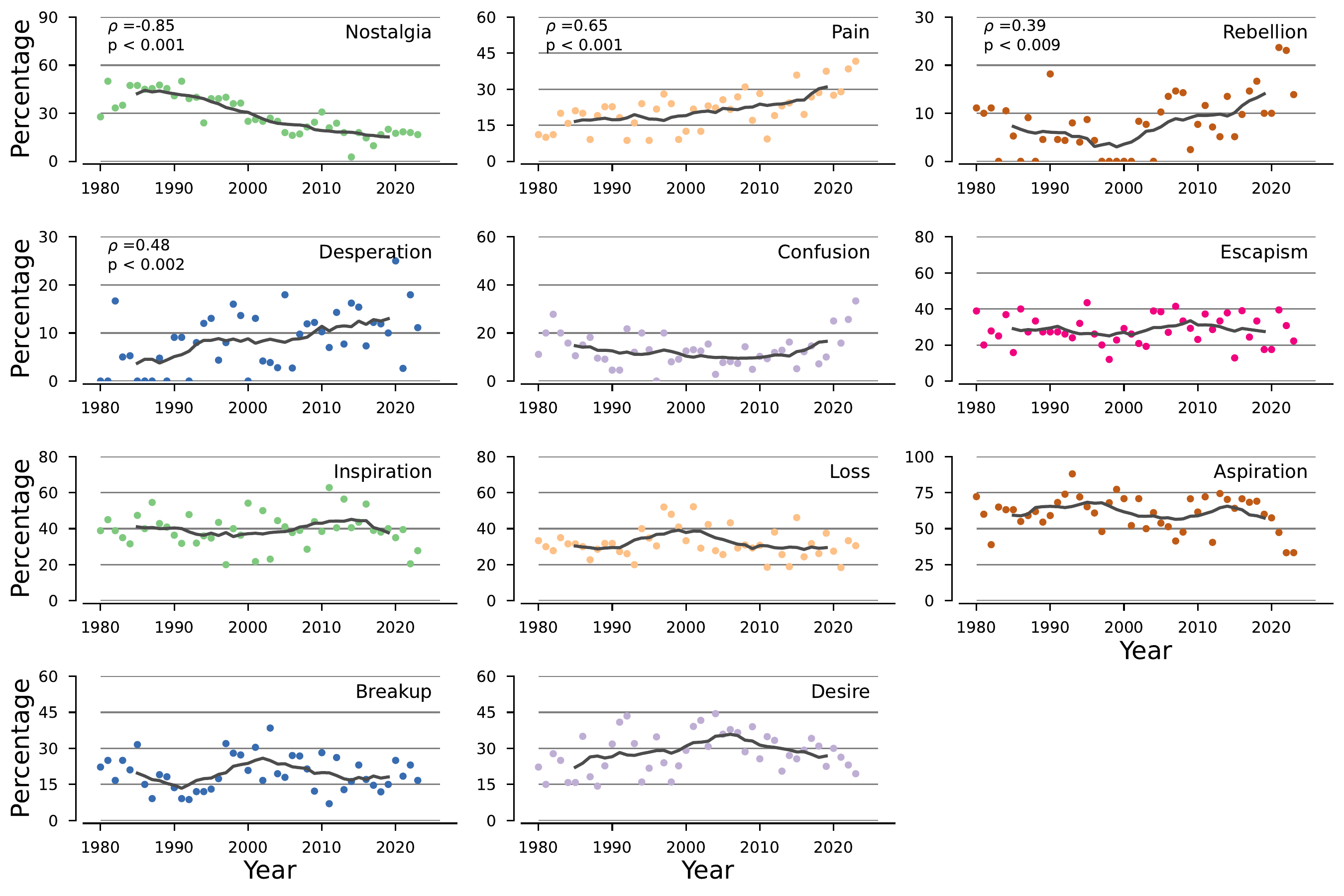}
    \caption{ \textbf{Temporal trends in theme usage by competing songs in the period 1980 to 2024. } By comparing with the previous figure, we see that some of the trends detected earlier where due to step changes around 1980. }
    \label{fig:topic_trends2}
\end{figure}

\begin{figure}[th!]
    \centering
    \includegraphics[width=0.48\linewidth]{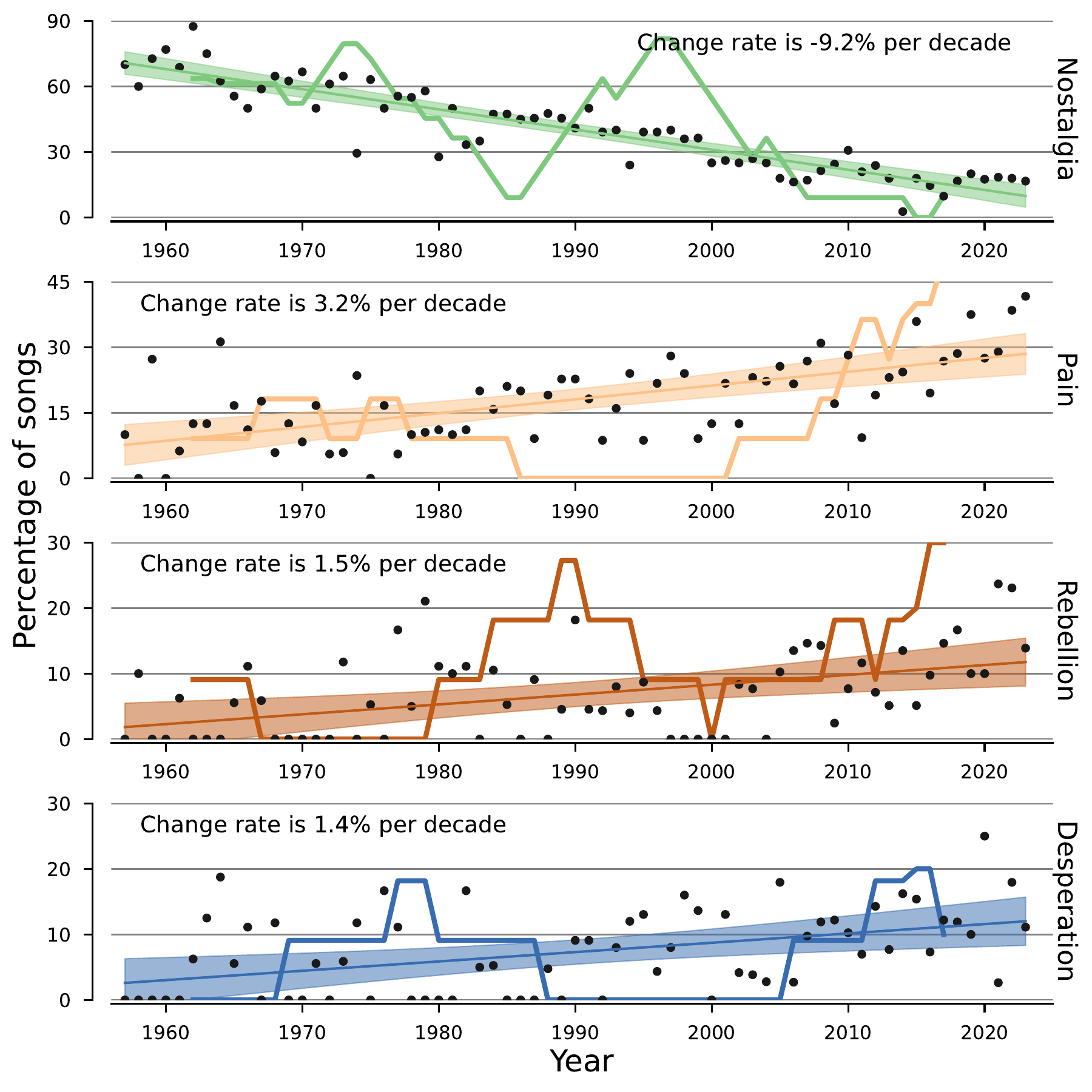}
    \includegraphics[width=.48\linewidth]{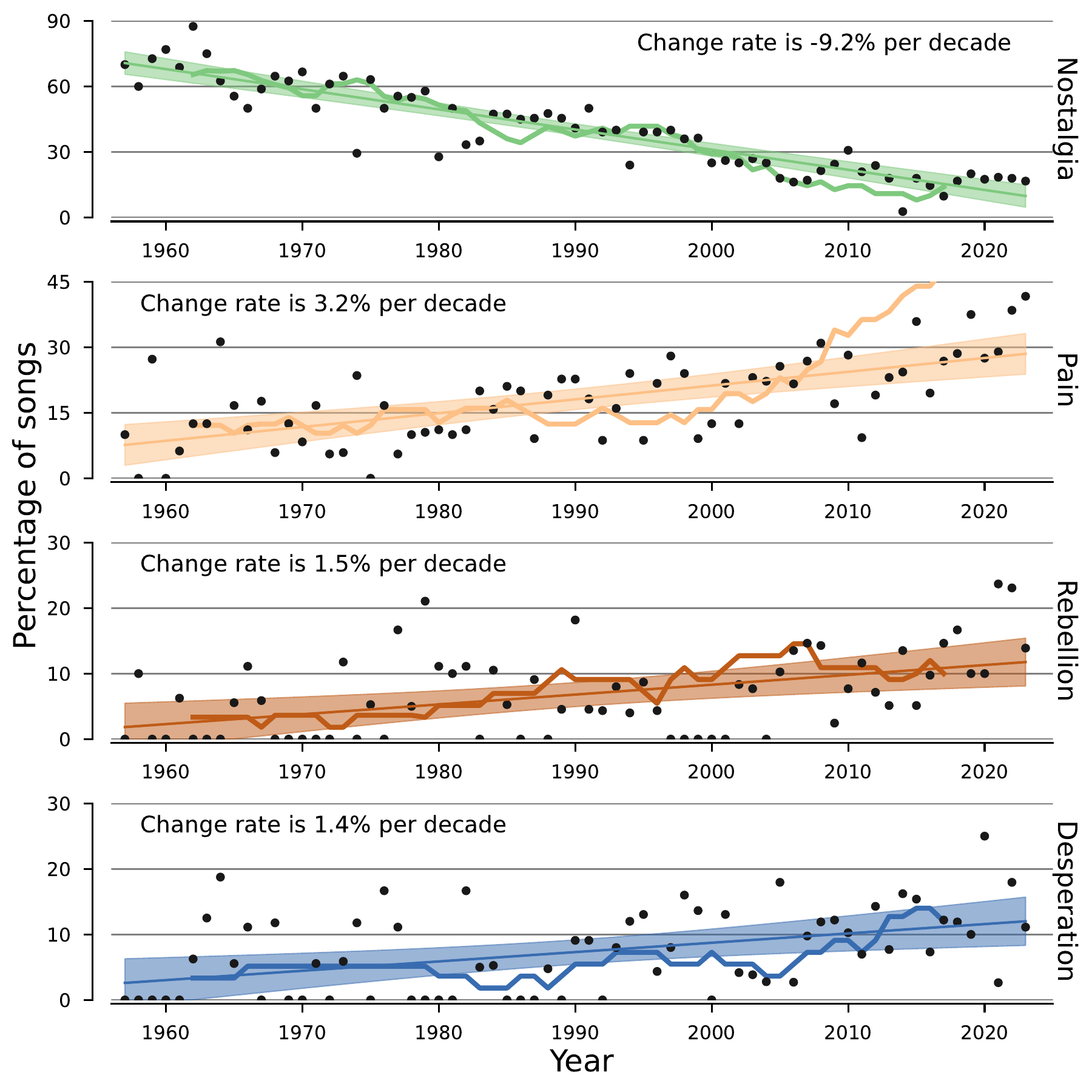}
    \caption{ {\bf Time evolution of the percentage representation of the four most informative song themes for a. winning songs and b. Top 5 songs.} 
    The black circles show average over all competing songs for a given year, the line and shaded regions show 99\% confidence intervals for the linear fit, and the colored lines show running 10-year running average of theme representation among Top 1 or Top 5 songs. 
    The trends observed for Top 3 songs (Fig.~\ref{fig:lyrics}b) are even more pronounced for the winning songs, with even a few additional patterns becoming apparent. For Top 5 songs, most of the trends observed fro Top 3 songs disappear. }
    \label{fig:topic_tops}
\end{figure}

\begin{figure}[th!]
    \centering
    \includegraphics[width=\linewidth]{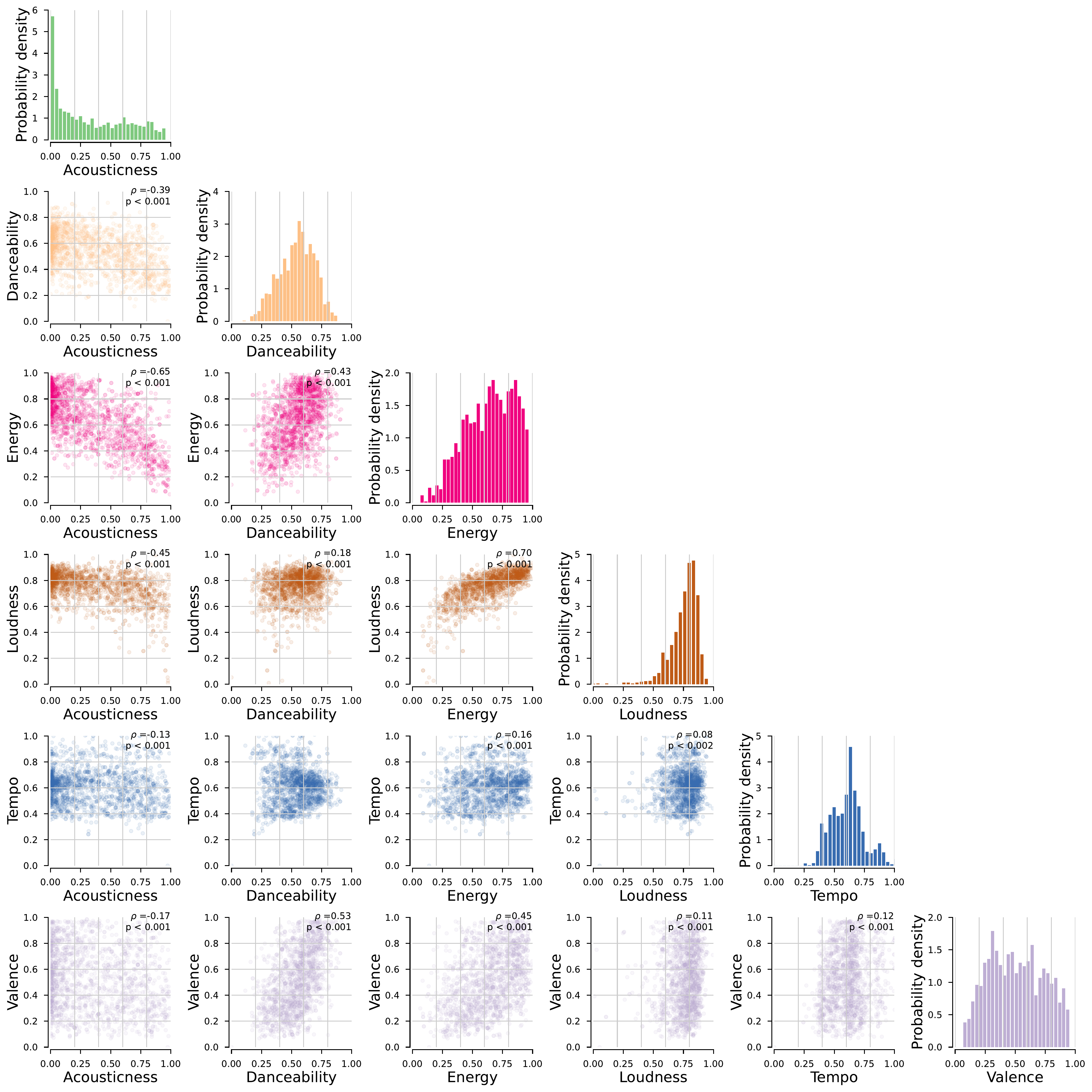}
    \caption{ \textbf{Characteristics of competing songs for all Contests between 1957 and 2024. } We use the Spotify API to extract six audio features from competing songs. To maintain a constant range for all variables, we normalize Tempo by dividing it by 200, and Loudness by the expression $1 + \ell / 30$.
    Along the diagonal, we show the distribution of feature scores for the six features.  The other plots show projections of the feature scores vector for each pair of two features.   We calculate Spearman's $\rho$ to estimate correlations between pairs of features and show results for the cases where the $p$-value of the estimated correlation coefficient is smaller than 0.05. We find that although Tempo is distributed over a limited range, it is statistically significantly correlated with several other features. However, the correlations are quite small in magnitude. 
    Not surprisingly, Energy and Loudness are strongly correlated, even though Loudness is distributed over a limited range. }
    \label{fig:spotify_features_all}
\end{figure}

\begin{figure}[ht!]
    \centering
    \includegraphics[width=0.8\linewidth]{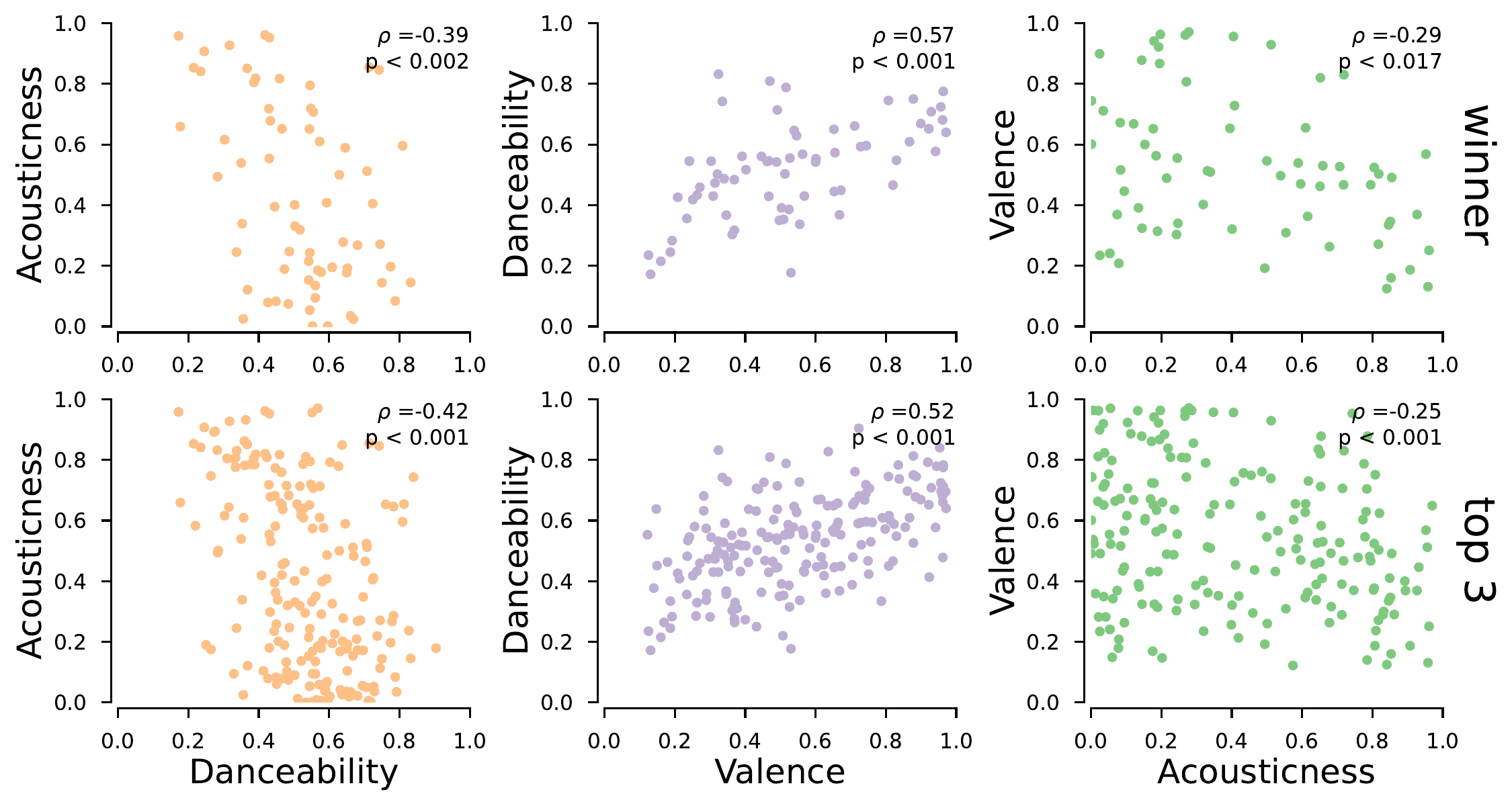}
    \caption{ \textbf{Cross-correlation of three audio features for the winning song and top three songs in the 1957--2023 Contests.}  Acousticness and danceability are strongly anti-correlated, as are acousticness and valence. Not surprisingly, danceability and valence are strongly correlated.  That is, danceable songs tend to be positive songs. }
    \label{fig:corr_spotify_features}
\end{figure}

\begin{figure}[ht!]
    \centering
    \includegraphics[width=0.8\linewidth]{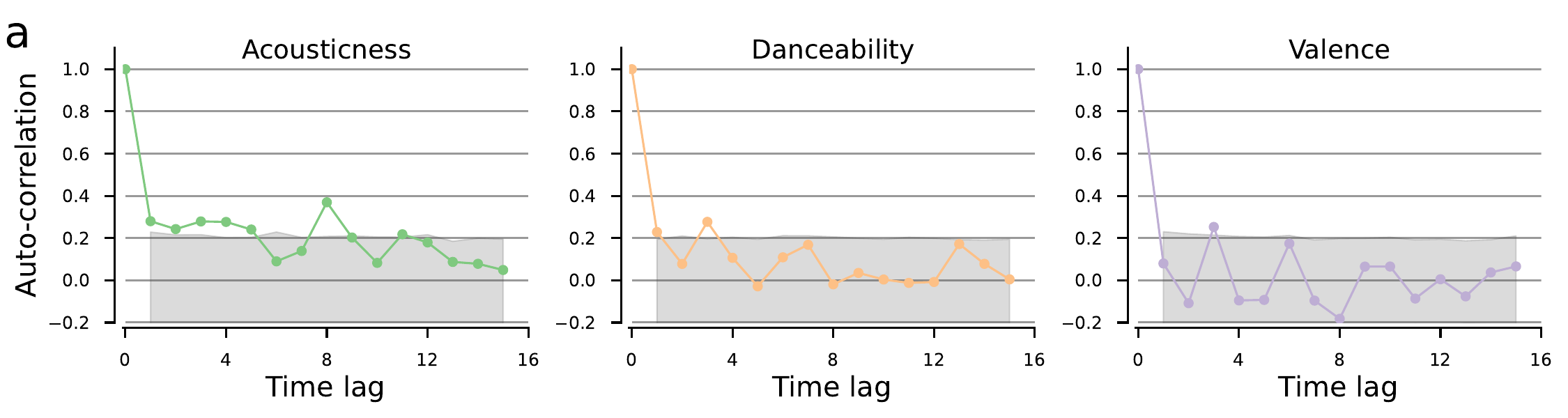}
    \includegraphics[width=0.8\linewidth]{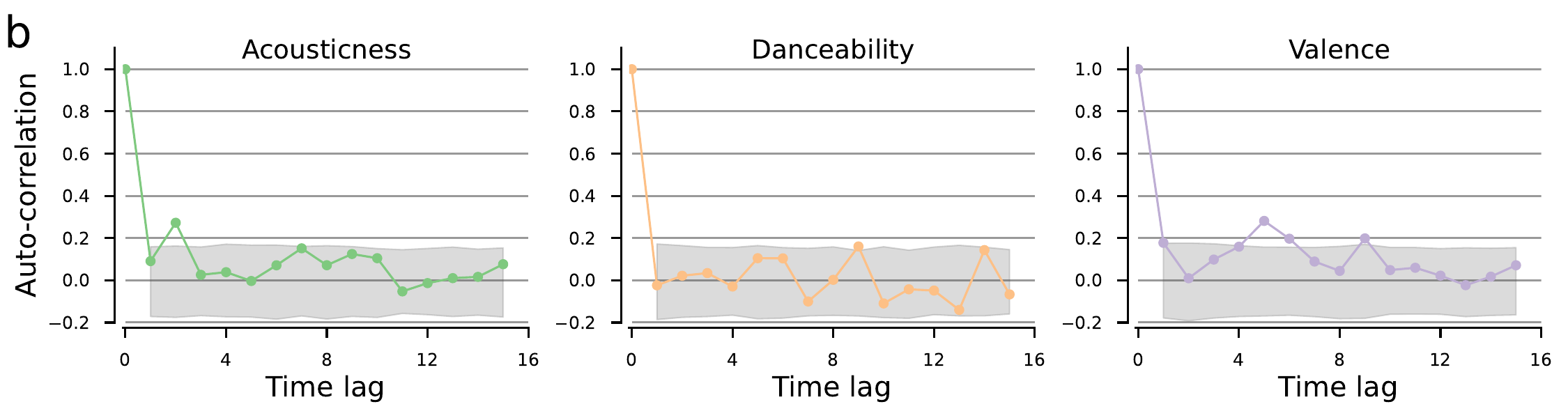}
    \caption{ \textbf{ Auto-correlation of audio features for 1957--1989 (a) and 1980--2024 (b).} For the earlier period, there are correlations over time for acousticness (as it is on a decaying trend) but not for the other features. For the later period, the temporal correlations practically disappear for all features. }
    \label{fig:feature_autocorrelation}
\end{figure}

\begin{figure}[th!]
    \centering
    \includegraphics[width=0.7\linewidth]{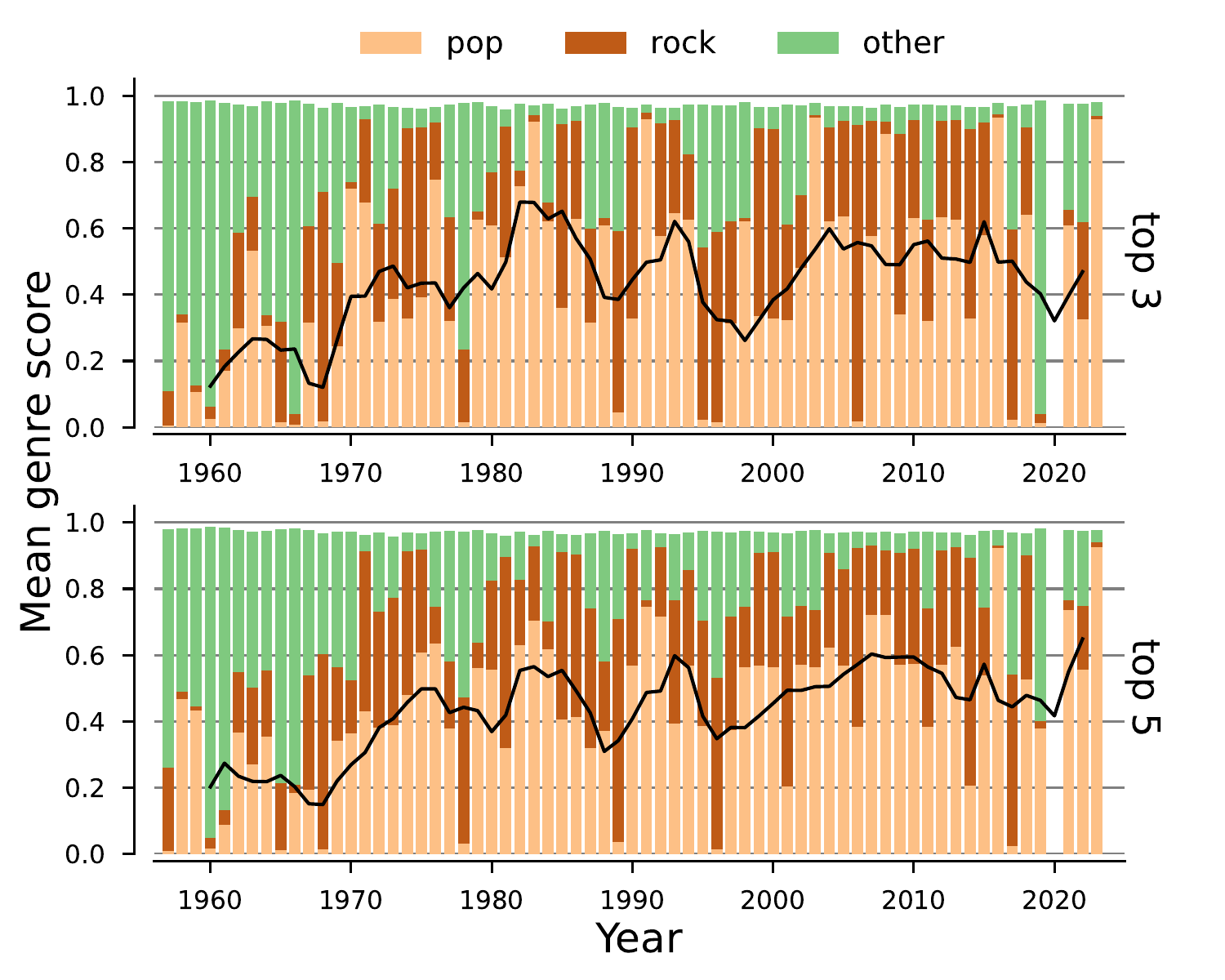}
    \caption{ \textbf{Mean annual genre score of top 3 and top 5 songs.}  
    Since about 1975 for top 3 songs and 1980 for top 5 songs, half of the top ranked songs have been pop songs.  These results suggest that, unlike with language, where English is the standard, there is more opportunity for ``bucking the trend" in song genre. } 

    \label{fig:music_genre_top3-5}
\end{figure}

\begin{figure}[ht!]
    \centering
    \includegraphics[width=\linewidth]{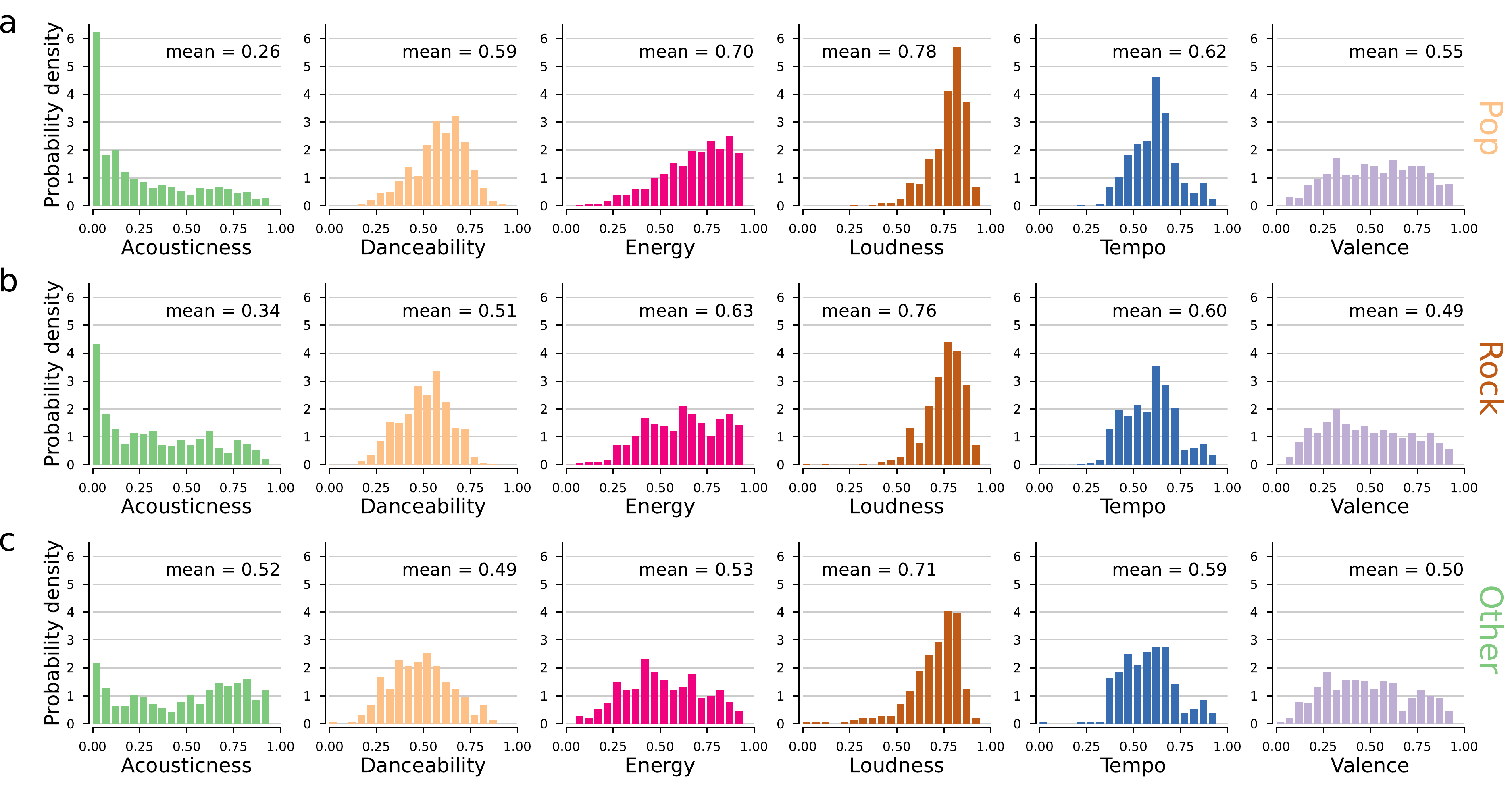}
    \caption{ \textbf{Song audio features change in expected ways with music genre.} 
    {\bf a.} Pop dominated songs (pop score greater than 0.66) have low acousticness, and high danceability, energy and tempo. Unlike the peaked distributions observed for those features, pop songs have a broad, perhaps bimodal, distribution of valences. 
    {\bf b.} Rock songs (rock score greater than 0.66) have broader distributions of audio feature values. The contrast with pop is the most marked for energy, perhaps due to rock ballads. 
    {\bf c.} Songs from other genres have a bimodal distribution of acousticnesses and tend to have lower danceability, energy, and valence.}  
    \label{fig:features_vs_genres}
\end{figure}

\begin{figure}[ht!]
    \centering
    \includegraphics[width=0.8\linewidth]{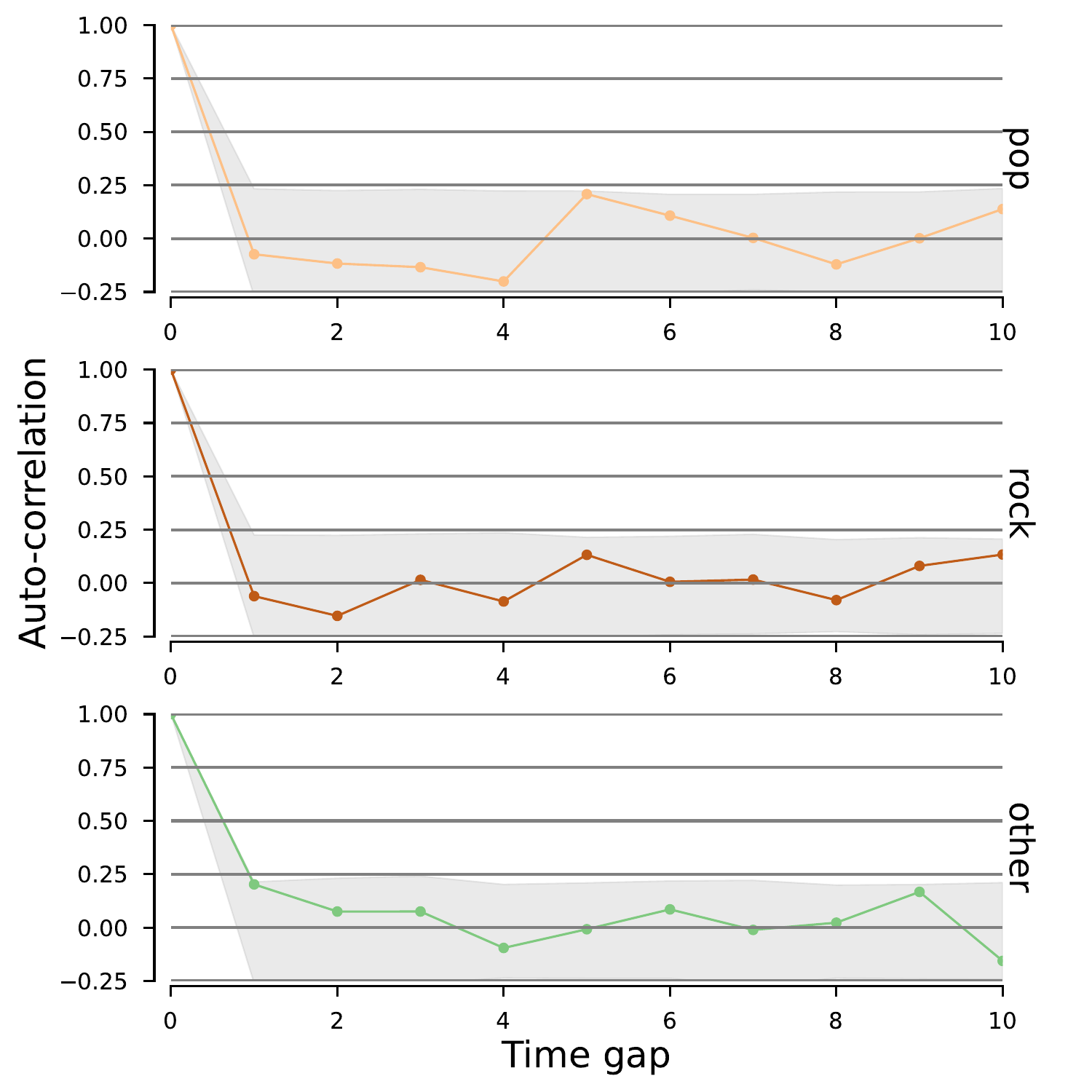}
    \caption{ \textbf{Auto-correlation of the genre score of the winning song for Pop, Rock and other genres.} Despite the apparent `up and down' nature of the data, the calculation of the auto-correlation shows that there are no significant temporal correlations of the genre, i.e., there is no predictability beyond the different winning probabilities of the genres.}
    \label{fig:genre_autocorrelation}
\end{figure}

\begin{figure}[ht!]
    \centering
    \includegraphics[width=0.9\linewidth]{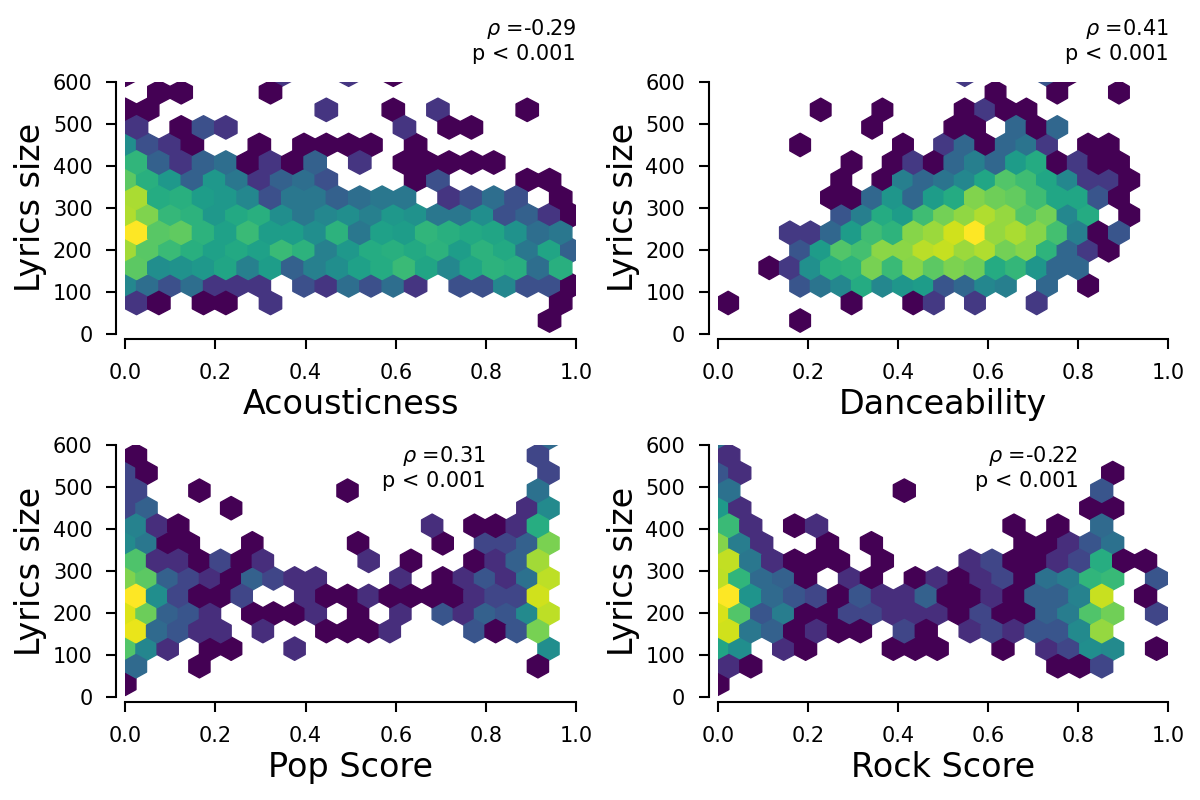}
    \caption{ \textbf{Correlation between lyrics size and genre score (top row) or audio feature (bottom row) for all competing songs.}  We use Spearman's rank correlation to quantify the degrees of correlation (and its statistical significance) between {\it number of words in lyrics\/} and other features. We find that as Acousticness or Rock score increase the number of words in a song's lyrics decreases.  In contrast, as Danceability or Pop score increase the number of words in song's lyrics also increases.}
    \label{fig:correlation_size_features_allyears}
\end{figure}

\begin{figure}[ht!]
    \centering    
    \includegraphics[width=0.9\linewidth]{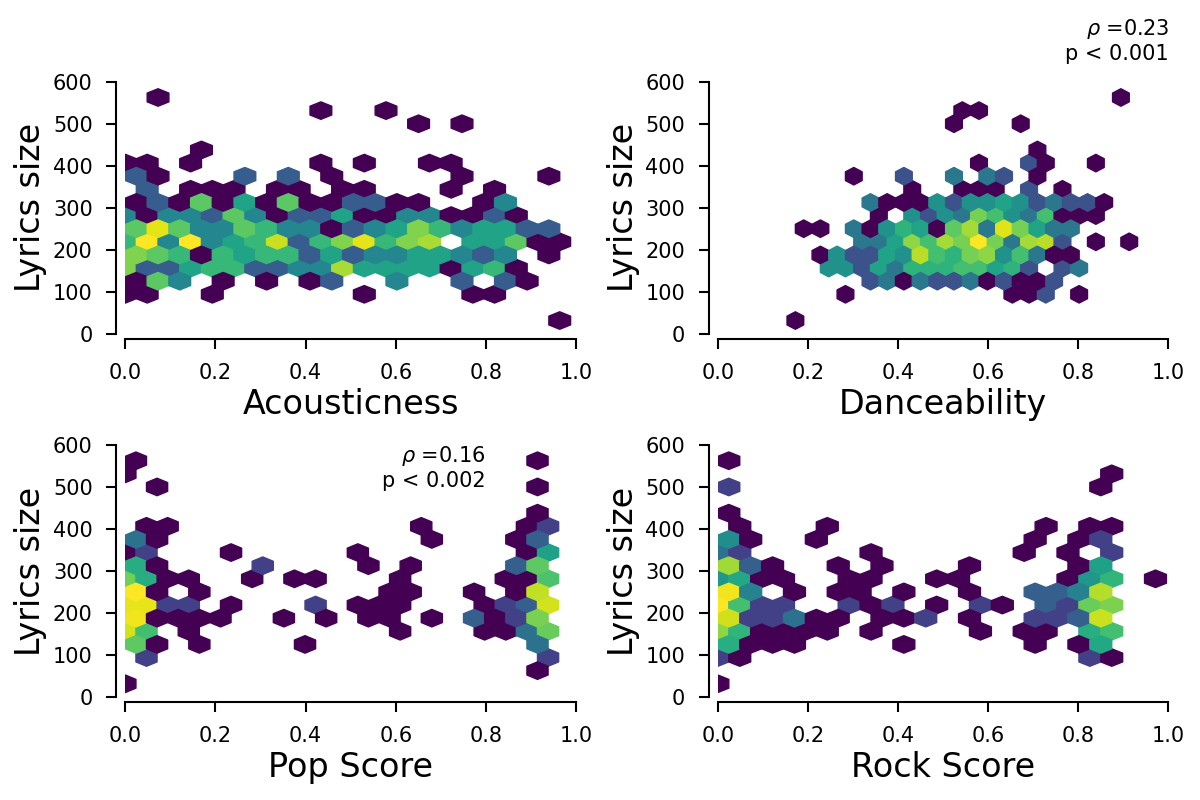}
    \caption{ \textbf{Correlation between lyrics size and genre score (top row) or audio feature (bottom row) for competing songs in the 1975 to 1998 Contests.}  We use Spearman's rank correlation to quantify the degrees of correlation (and its statistical significance) between {\it number of words in lyrics\/} and other features. We find that as Danceability or Pop score increase the number of words in song's lyrics also increases. In contrast, we find no significant correlation between Acousticness or Rock score and the number of words in a song's lyrics decreases.}
    \label{fig:correlation_size_features_early}
\end{figure}

\begin{figure}[ht!]
    \centering    
    \includegraphics[width=0.9\linewidth]{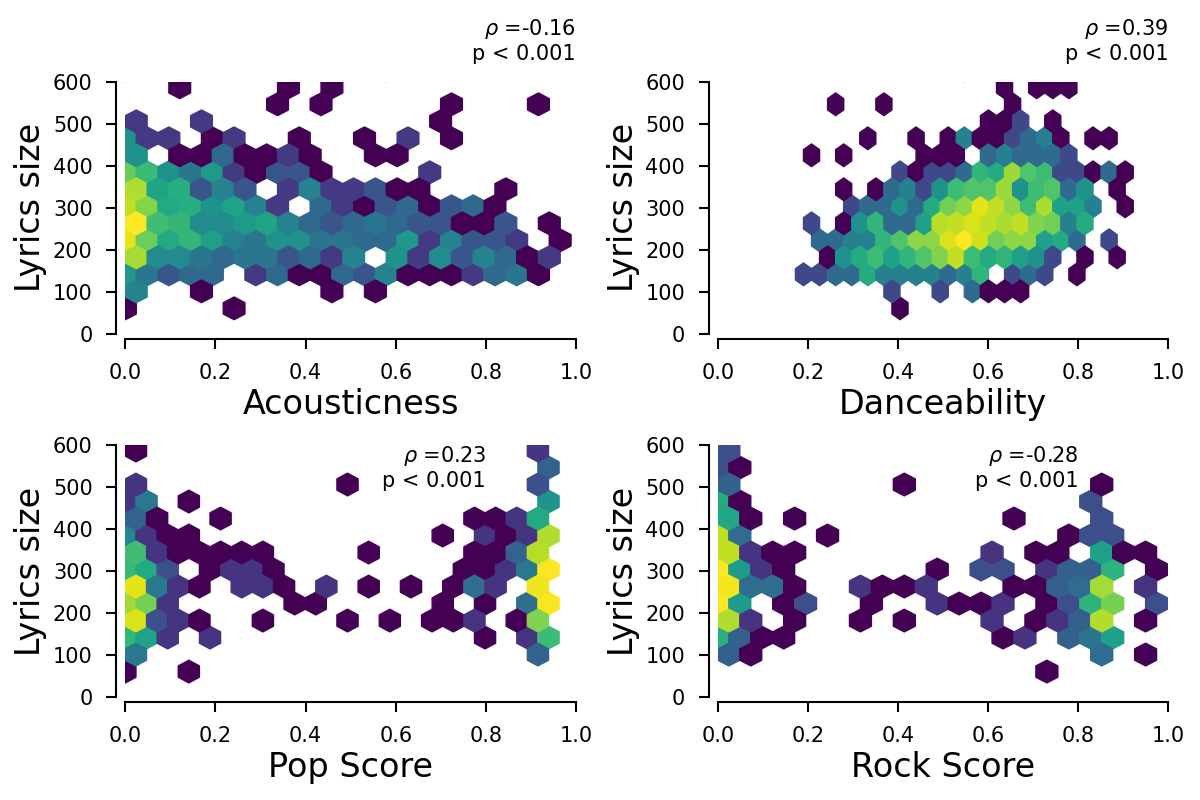}
    \caption{ \textbf{Correlation between lyrics size and genre score (top row) or audio feature (bottom row) for competing songs in the 1999 to 2024 Contests.}  We use Spearman's rank correlation to quantify the degrees of correlation (and its statistical significance) between {\it number of words in lyrics\/} and other features. We find that as Acousticness or Rock score increase the number of words in a song's lyrics decreases.  In contrast, as Danceability or Pop score increase the number of words in song's lyrics also increases.}
    \label{correlations_lyrics_size_audio_features_1999_2024}
\end{figure}

\begin{figure}[ht!]
    \centering
    \includegraphics[width=0.9\linewidth]{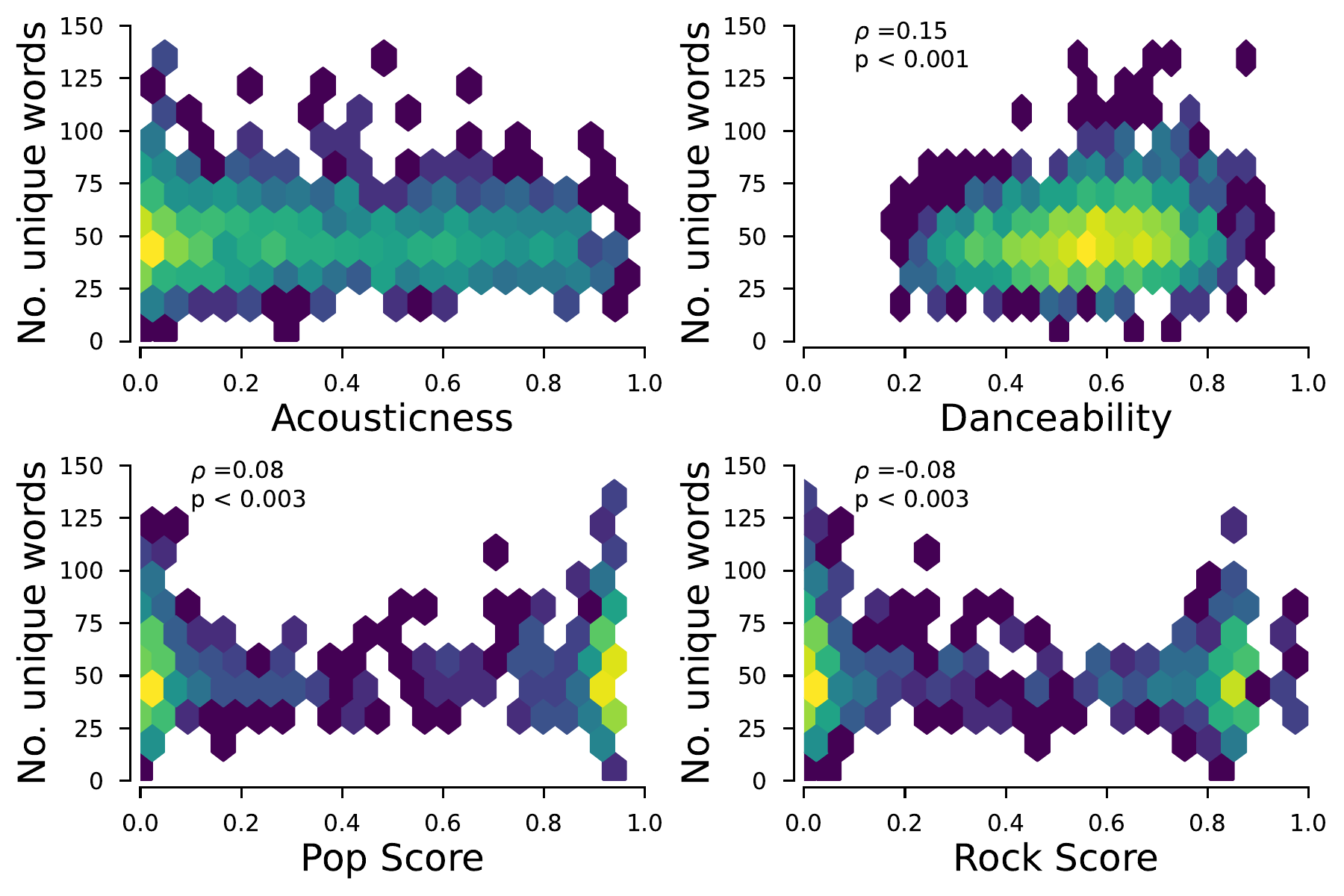}
    \caption{ \textbf{Correlation between lyrics size and genre score (top row) or audio feature (bottom row) for competing songs in the 1975 to 2024 Contests.}  We use Spearman's rank correlation to quantify the degrees of correlation (and its statistical significance) between {\it number of unique words\/} in lyrics and other features. We find that as Rock score increase the number of unique words in a song's lyrics decreases.  In contrast, as Danceability or Pop score increase the number of unique words in the song's lyrics also increases.}
    \label{fig:correlations_number_unique_words_audio_features_1975_2024}
\end{figure}

\begin{figure}[ht!]
    \centering    
    \includegraphics[width=0.9\linewidth]{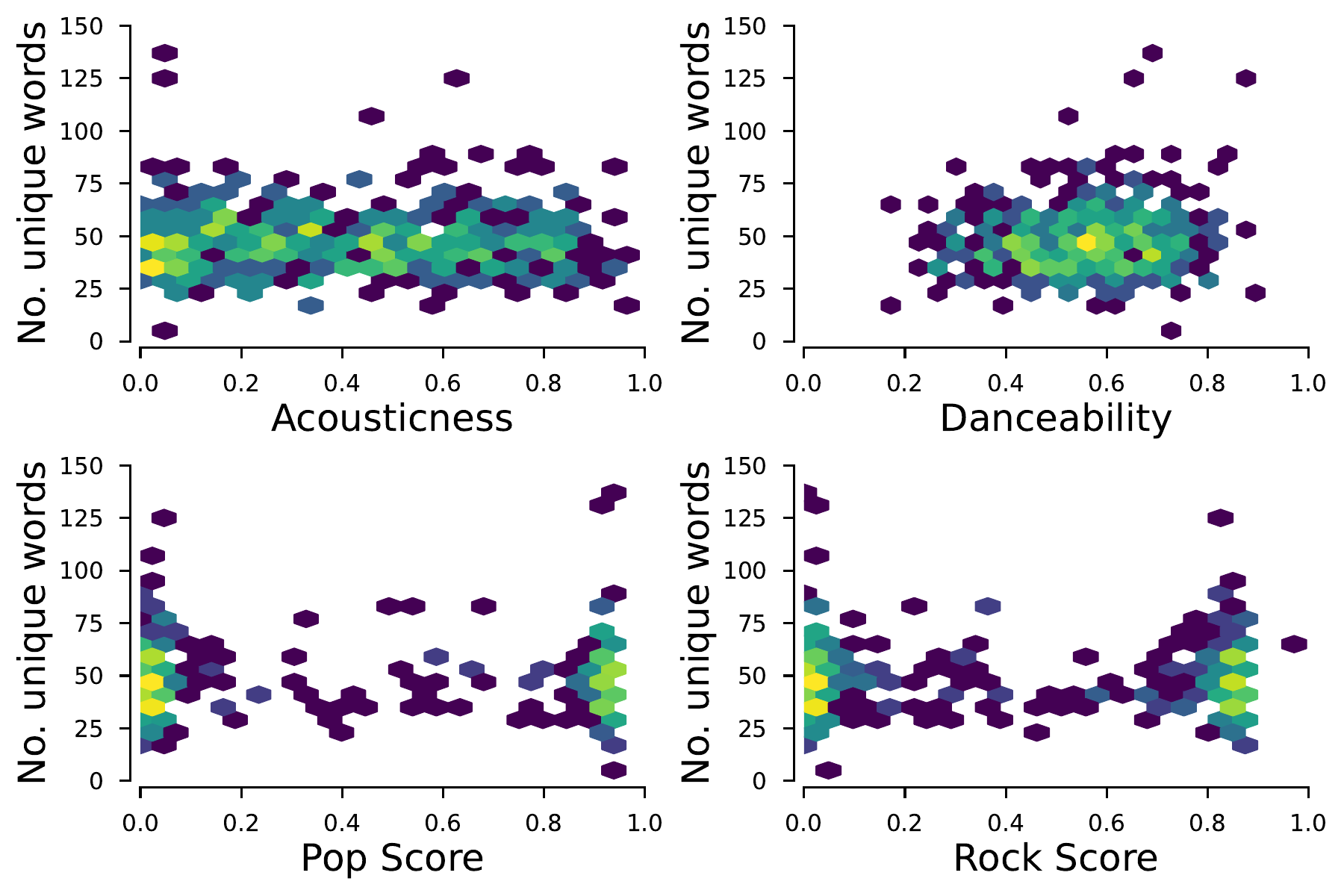}
    \caption{ \textbf{Correlation between lyrics size and genre score (top row) or audio feature (bottom row) for competing songs in the 1975 to 1998 Contests.}  We use Spearman's rank correlation to quantify the degrees of correlation (and its statistical significance) between {\it number of unique words\/} in lyrics and other features. We find no statistically significant correlations.}
    \label{correlations_number_unique_words_audio_features_1975_1998}
\end{figure}

\begin{figure}[ht!]
    \centering    
    \includegraphics[width=0.9\linewidth]{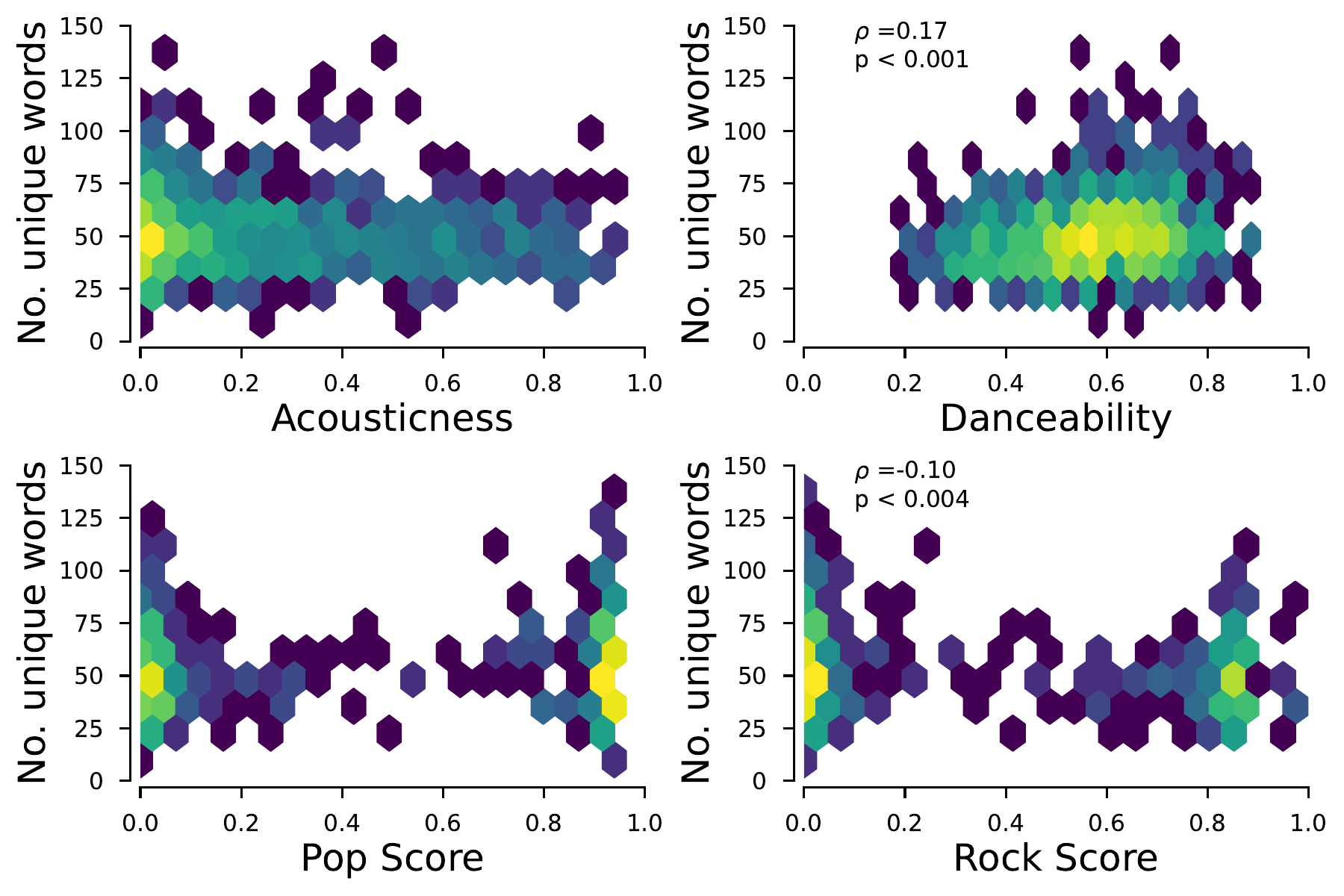}
    \caption{ \textbf{Correlation between lyrics size and genre score (top row) or audio feature (bottom row) for competing songs in the 1999 to 2024 Contests.}  We use Spearman's rank correlation to quantify the degrees of correlation (and its statistical significance) between {\it number of unique words\/} in lyrics and other features. We find that as Rock score increase the number of unique words in a song's lyrics decreases.  In contrast, as Danceability increases the number of unique words in the song's lyrics also increases.}
    \label{correlations_number_unique_words_audio_features_1999_2024}
\end{figure}

\begin{figure}[ht!]
    \centering    
    \includegraphics[width=0.48\linewidth]{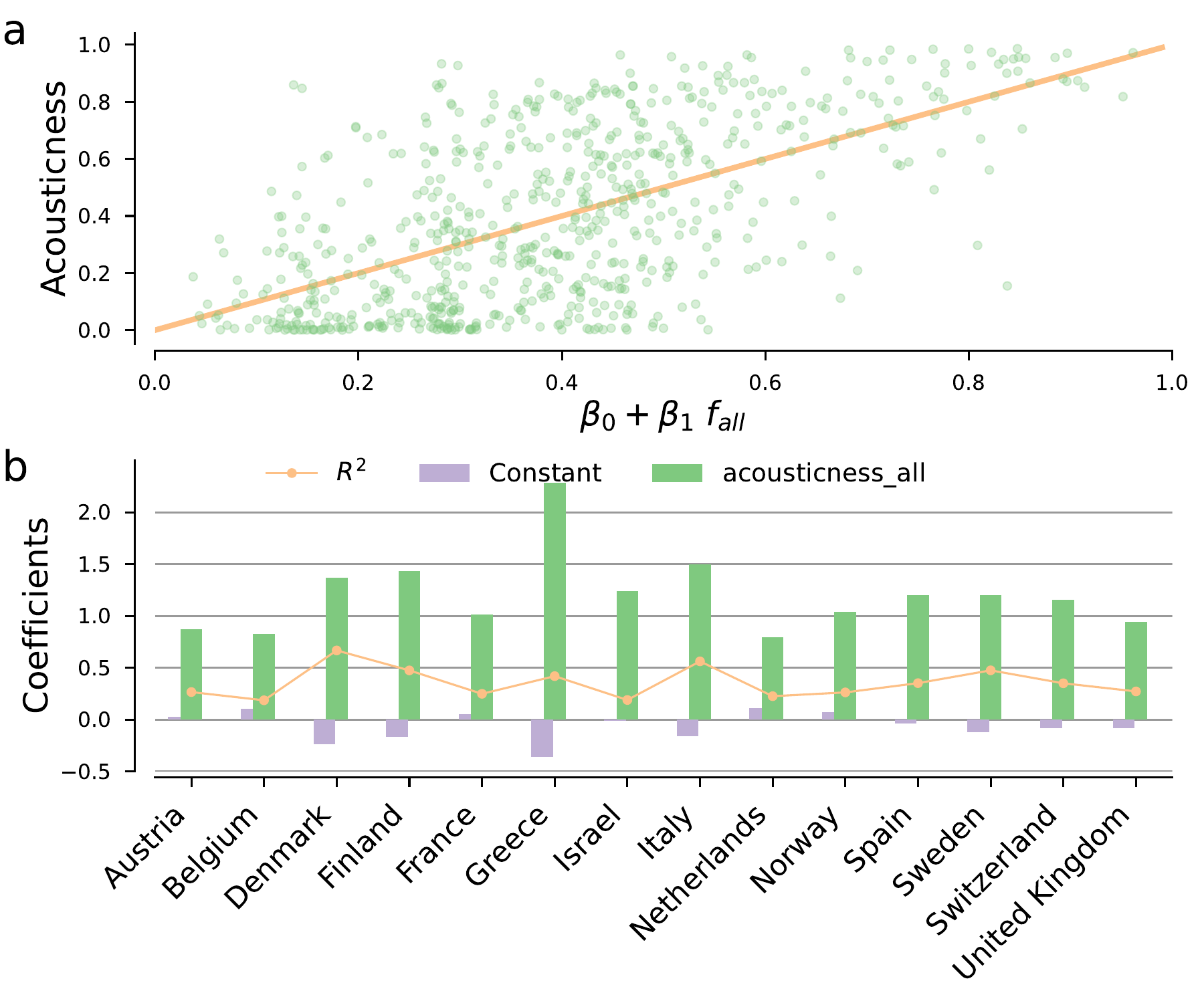}
    \includegraphics[width=0.48\linewidth]{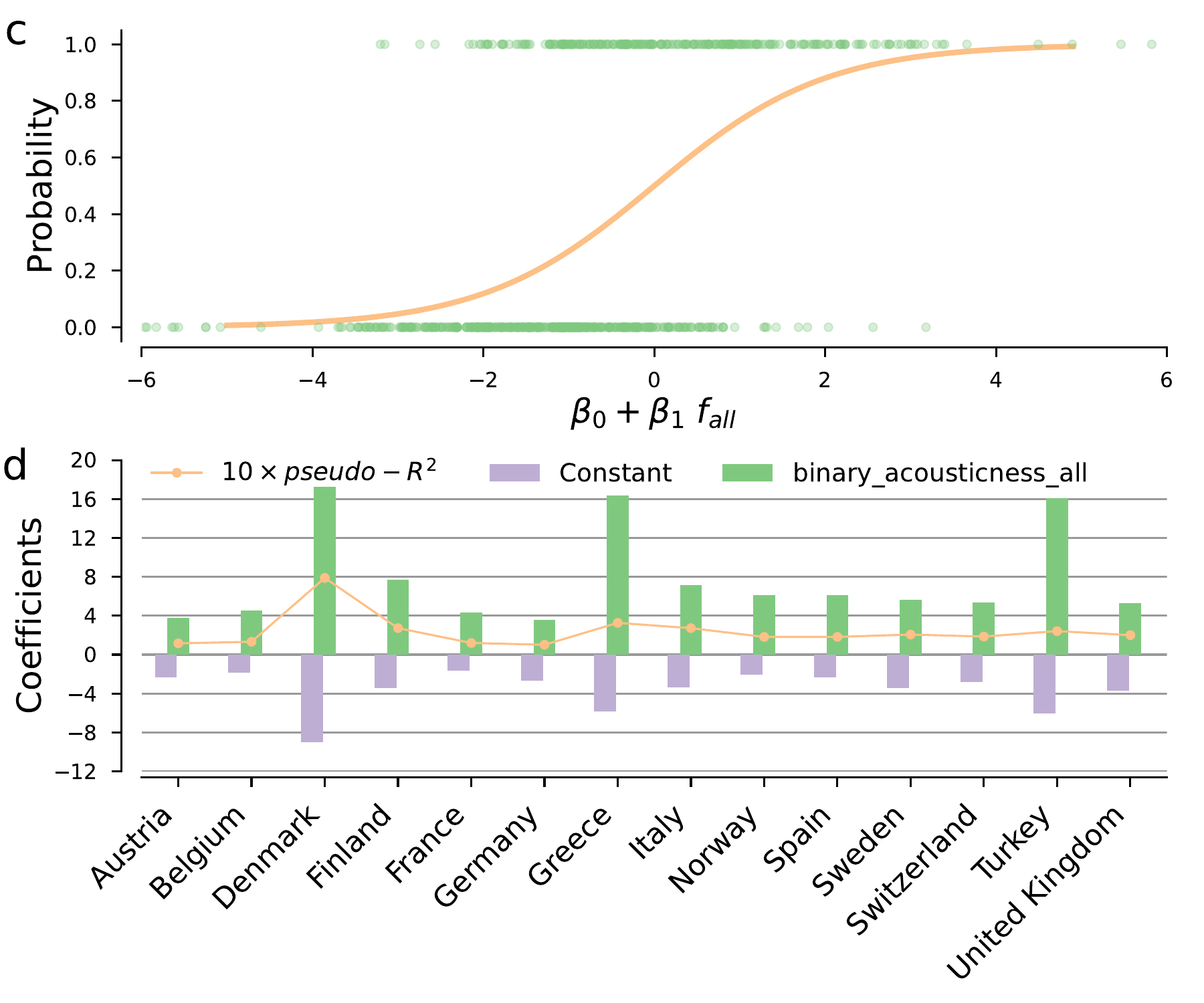}
    \caption{ \textbf{Country learning of interest in acousticness from competing songs.} 
    {\bf a.} Linear regression of acousticness of a country's song in year $y$ against the average acousticness of {\it all\/} other countries in the previous 3 years.
    {\bf b.} Regression coefficients for the 14 countries for which the linear model is statistically significant and $R^2$ of the linear fit. 
    {\bf c.} Logistic regression of a country's song being classified as acoustic --- i.e., having an acousticness score greater than 0.5 --- in year $y$ to the average fraction of songs classified as acoustic in the previous 3 years.
    {\bf d.} Regression coefficients for the 14 countries for which the logistic model is statistically significant and  pseudo-$R^2$ of the fit. }
    \label{fig:regression_acousticness}
\end{figure}

\clearpage
\newpage
\subsection*{Prompt GPT-4o for Lyrics Topic Modelling}\label{sec:prompt}

The following code is the specific message that is passed as a parameter to the GPT-4o API for the task of modeling the topic of the lyrics. The topics and their associated keywords were taken from the work of Henard and Rossetti~\cite{henard2014all}, who defined various topics based on emotional and thematic categories in song lyrics. The query to the GPT-4o model was executed on the 9th of December 2024.

\begin{lstlisting}[frame=single, framesep=10pt, basicstyle=\tiny\ttfamily, breaklines=true, language=Python]

messages = [
    {"role": "system", "content": "You are a helpful assistant that classifies song lyrics into topics."},
    {"role": "user", "content": """
    Given the following song title and lyrics, assign up to 3 topics.
    Use the topics in this dictionary. The topics are the keys of the dictionary. Each of these topics has a set of descriptors and key influential words.
    
    {"Loss": {"Associated Descriptors": "Sadness; Heartache; Losing Love; Loneliness; Unrealized Romance",
      "Key Influential Words": "Heart; Love; Apart; Lonely; Leaving; Feeling"},
     "Desire": {"Associated Descriptors": "Longing; Wanting; Anticipation; Lust; Coveting",
      "Key Influential Words": "Move; Tonight; Baby; Body; Trying; Dance"},
     "Aspiration": {"Associated Descriptors": "Dreams; Other Worldly; Longing; Anticipation; Hope",
      "Key Influential Words": "Dream; Heaven; Tonight; Moon; Move; World"},
     "Breakup": {"Associated Descriptors": "Out of Love; Loss; Separation; Good-bye; Person-specific",
      "Key Influential Words": "Waiting; Time; Leaving; Tear; Good-bye; Lonely"},
     "Pain": {"Associated Descriptors": "Mixed Emotions; In the Moment; Sadness; Change; Good-bye",
      "Key Influential Words": "Bad; Good; Good-bye; Kiss; Moment; Woman"},
     "Inspiration": {"Associated Descriptors": "Optimism; Energetic; Vibrancy; Dancing; Connectedness",
      "Key Influential Words": "Life; Everything; Music; Good; World; Dancing"},
     "Nostalgia": {"Associated Descriptors": "Romantic; Oneness; Togetherness; Dreamy; Idealistic",
      "Key Influential Words": "Feeling; Morning; Woman; Heart; Time; Heaven"},
     "Rebellion": {"Associated Descriptors": ["Rebellious", "Counter Culture", "Rock 'n' Roll", "Music"],
      "Key Influential Words": ["Favorite", "Sunday", "Playing", "City", "Rock", "Radio"]},
     "Jaded": {"Associated Descriptors": ["Reflective", "Jaded", "Cynicism", "Trapped", "Contemplative"],
      "Key Influential Words": ["State", "Favorite", "Mirror", "Sunday", "Heavy", "Jet"]},
     "Desperation": {"Associated Descriptors": ["Helpless",
     "Trapped", "Desperate", "Cornered", "Coveting"],
      "Key Influential Words": ["Corner", "Throw", "Line", "Guess", "Promise", "Stupid"]},
     "Escapism": {"Associated Descriptors": ["Fantasy", "Escape", "Ecstasy", "Love as a Drug", "Sex"],
      "Key Influential Words": ["Deal", "Ecstasy", "Flying", "Inside", "Question", "Grooving"]},
     "Confusion": {"Associated Descriptors": ["Confused", "Distant Memories", "Pointless", "Secrets"],
      "Key Influential Words": ["Memory", "Suitcase", "Circle", "View", "Secret", "Magic"]}}
      
    Use contextual reasoning to determine the most appropriate topics. Help the contextual reasoning with the information presented in the Associated Descriptors and Key Influential Words.
    Respond ONLY with the relevant topics separated by a comma. Don't include any description of why you included those topics. The topics must be sorted by relevance.
    """+f"""
    Title: {title}
    Lyrics: {lyrics}

    Response:"""
]
\end{lstlisting}


\clearpage
\newpage
\subsection*{Decanal changes in song characteristics}

To further validate the audio features produced by the Spotify API, we explored whether their values could be used to identify the decade in which a song was created. We trained five classifiers --- Logistic Regression, Gaussian Naive Bayes, Random Forest, Support Vector Machine, and Xgboost --- using the following features: Danceability, Energy, Loudness, Acousticness, Valence, Tempo, and genre represented using one-hot encoding. We optimized the hyper-parameters for each classifier via grid search. We then evaluated their performance according to the F1 measure (harmonic mean of precision and recall) using 10-fold cross-validation. 

\begin{figure}[b!]
    \centering
    \includegraphics[width=0.7\linewidth]{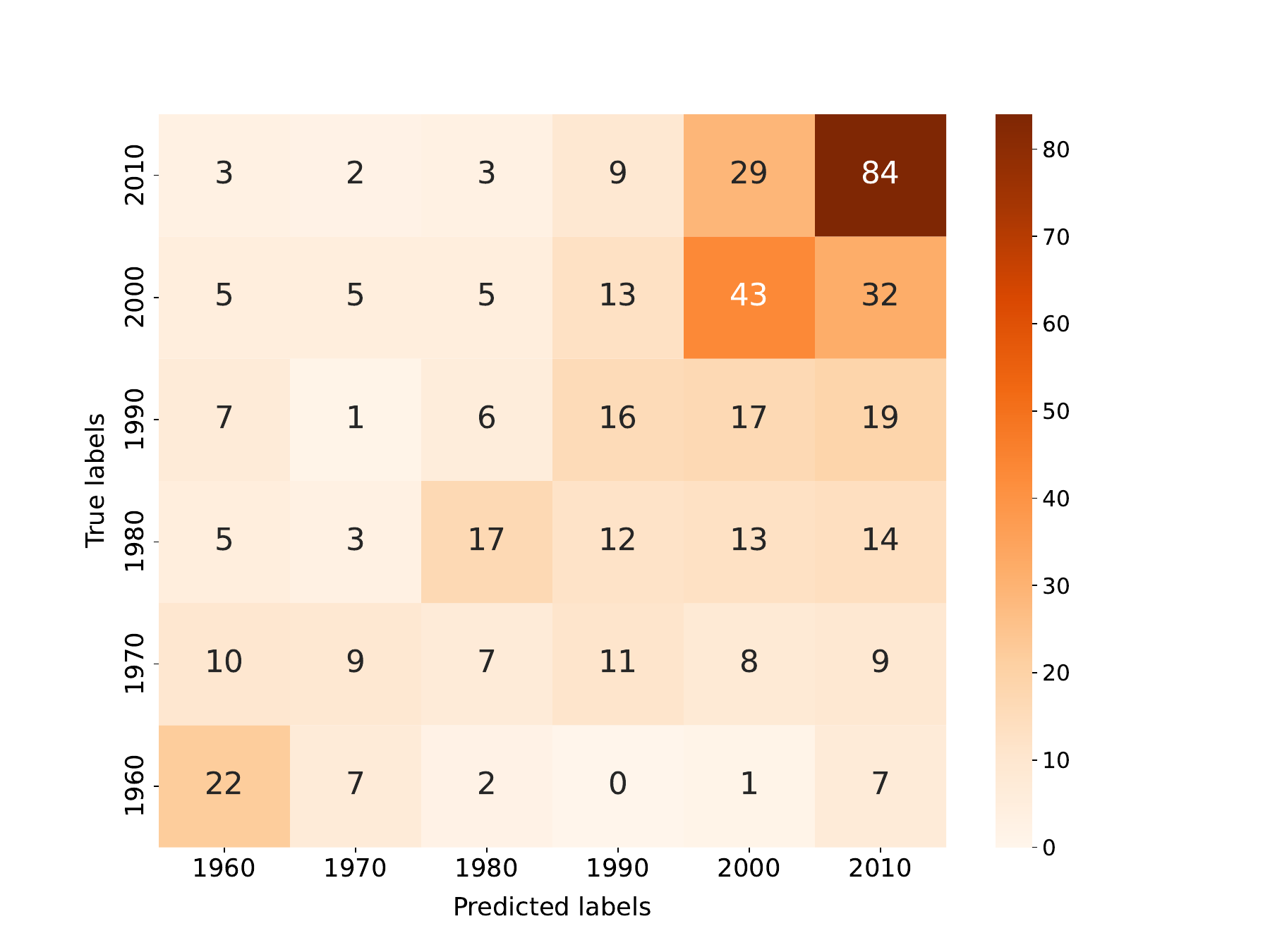}
    \caption{Confusion matrix of true labels versus predicted labels for the task of predicting the decade of a song given its audio features.}
    \label{fig:predict_decade}
\end{figure}

We used a Random Forest Classifier and determined the optimal hyperparameters through a grid search: a maximum depth of 10, 100 estimators, and a minimum of 10 samples per split. Our model achieved an average weighted F1 score of $0.41$, with the 1960s and 1990s being the most challenging decades to classify. This performance is significantly higher than the expectation for a random classifier applied to an unbalanced eight-class task classification (F1 $= 0.18$). Figure~\ref{fig:predict_decade} shows the confusion matrix of true labels versus predicted labels for the Random Forest classifier. Reassuringly, we observed that mis-classification is most frequent within one decade of the correct time.
    
\end{document}